\documentclass[final,preprint,5p,number,sort&compress,reversenotenum]{elsarticle}

\usepackage{amssymb}
\usepackage{amsmath}
\usepackage[final, hidelinks]{hyperref}
\usepackage[obeyFinal]{todonotes}
\setuptodonotes{inline, size=\small, author=NOAUTHOR}
\usepackage{lineno}
\usepackage{orcidlink}
\usepackage{ifdraft}

\graphicspath{{Figures/}}
\journal{Nuclear Inst. and Methods in Physics Research, A}

\makeatletter
\def\ps@pprintTitle{%
  \let\@oddhead\@empty
  \let\@evenhead\@empty
  \def\@oddfoot{\footnotesize\itshape
       Preprint accepted by \@journal\hfill\@date}%
  \let\@evenfoot\@oddfoot
}
\makeatother

\begin{document}

\begin{frontmatter}

  \title{Machine-learning correction for the calorimeter saturation of cosmic-ray ions with the Dark Matter Particle Explorer: towards the PeV scale}

  \author[dpnc]{Andrea Serpolla\,\orcidlink{0000-0002-4122-6298}\corref{cor}}
  \cortext[cor]{Corresponding author}
  \ead{andrea.serpolla@cern.ch}

  \author[dpnc]{Andrii Tykhonov}
  \author[dpnc]{Paul Coppin}
  \author[dpnc]{Manbing Li}
  \author[dpnc]{Andrii Kotenko}
  \author[dpnc]{Enzo Putti-Garcia}
  \author[dpnc]{Hugo Valentin Boutin}
  \author[issi]{Mikhail Stolpovskiy}
  \author[epfl]{Jennifer Maria Frieden}
  \author[epfl]{Chiara Perrina}
  \author[dpnc]{Xin Wu}

  \affiliation[dpnc]{
    organization={Département de physique nucléaire et corpusculaire (DPNC), Université de Genève (UniGE)},
    addressline={Quai Ernest-Ansermet 24},
    city={Geneva},
    postcode={CH-1205},
    country={Switzerland}}

  \affiliation[epfl]{
    organization={Institute of Physics, Ecole Polytechnique Fédérale de Lausanne (EPFL)},
    city={Lausanne},
    postcode={CH-1015},
    country={Switzerland}}

  \affiliation[issi]{
    organization={International Space Science Institute (ISSI)},
    addressline={Hallerstrasse 6},
    city={Bern},
    postcode={CH-3012},
    country={Switzerland}}

  \begin{abstract}
    The Dark MAtter Particle Explorer (DAMPE) instrument is a space-borne cosmic-ray detector, capable of measuring ion fluxes up to $\sim$500~TeV/n.
    This energy scale is made accessible through its calorimeter, which is the deepest currently operating in orbit.
    Saturation of the calorimeter readout channels starts occurring above $\sim$100~TeV of incident energy, and can significantly affect the primary energy reconstruction.
    Different techniques---analytical and machine-learning based---were developed to tackle this issue, focusing on the recovery of single-bar deposits, up to several hundreds of TeV.
    In this work, a new machine-learning technique is presented, which benefits from a unique model to correct the total deposited energy in DAMPE calorimeter.
    The described method is able to generalise its corrections for different ions and extend the maximum detectable incident energy to the PeV scale.
    This work is a continuation of the results presented in \cite{DAMPE_saturation_ML_2022}.
  \end{abstract}


\end{frontmatter}

\ifoptiondraft{\linenumbers}{}

\section{Introduction}\label{sec:intro}

Galactic cosmic rays (GCRs) are accelerated high-energy particles wandering in our galaxy.
Their origin and interaction with the interstellar medium (ISM) are crucial topics in astrophysics \cite{CR_physics}.
One of the main challenges in GCR studies is the accurate measurement of their composition and energy, which are essential to probe their origin in our galaxy and propagation in the ISM.
More specifically, a key open question is the behaviour of heavy nuclei fluxes at the 100~TeV to PeV scale; this work is a necessary and important step to achieve such goal.

The DArk Matter Particle Explorer (DAMPE) is a space-based detector operating since its launch in December 2015 \cite{DAMPE_mission}.
DAMPE is capable of detecting CR nuclei, electrons/positrons and $\gamma$-rays, thanks to its four sub-detectors: a plastic scintillator (PSD) \cite{DAMPE_PSD}, a silicon-tungsten tracker-converter (STK) \cite{DAMPE_STK}, a bismuth germanium oxide (BGO) electromagnetic calorimeter \cite{DAMPE_BGO}, and a neutron detector (NUD) \cite{DAMPE_NUD}.
The DAMPE instrument allows to detect and study $\gamma$-rays and electrons/positrons from $\sim$5~GeV up to several TeVs \cite{DAMPE_gamma_2022, DAMPE_electron_2017}, and nuclei from $\sim$50~GeV up to several hundreds of TeVs \cite{DAMPE_pHe_2024, DAMPE_boron_2025, DAMPE_BC_BO_ratios_2022}.
DAMPE stands out over other space-based instruments because of its calorimeter, the deepest currently in orbit with its 32 radiation lengths, or 1.6 nuclear interaction lengths.
Above $\sim$100~TeV of kinetic energy, the readout channels can saturate and a significant fraction of the deposited particle energy can be lost.

Previous studies showed the possibility of recovering the lost energy due to saturation using analytical \cite{DAMPE_saturation_analytical}, or machine-learning (ML) techniques \cite{DAMPE_saturation_ML_2022}.
However, these methods start losing accuracy and precision for heavy nuclei, and for incident energies above $\sim$500~TeV.
Therefore, the development of techniques that can help correcting for saturation at higher energies and for heavy nuclei is crucial for the measurement of GCRs spectra beyond the multi-TeV scale.
In this work, we present a ML method to correct the total deposited energy in DAMPE calorimeter for saturation up to few PeVs of primary energy, that can be applied to light ions (e.g., protons, helium) as well as on heavier ones (e.g., carbon, oxygen, iron).

\section{The BGO calorimeter of DAMPE}\label{sec:calorimeter}

\begin{figure}[htb]
  \centering
  \includegraphics[width=\ifoptiondraft{0.5\linewidth}{0.8\linewidth}, draft=false]{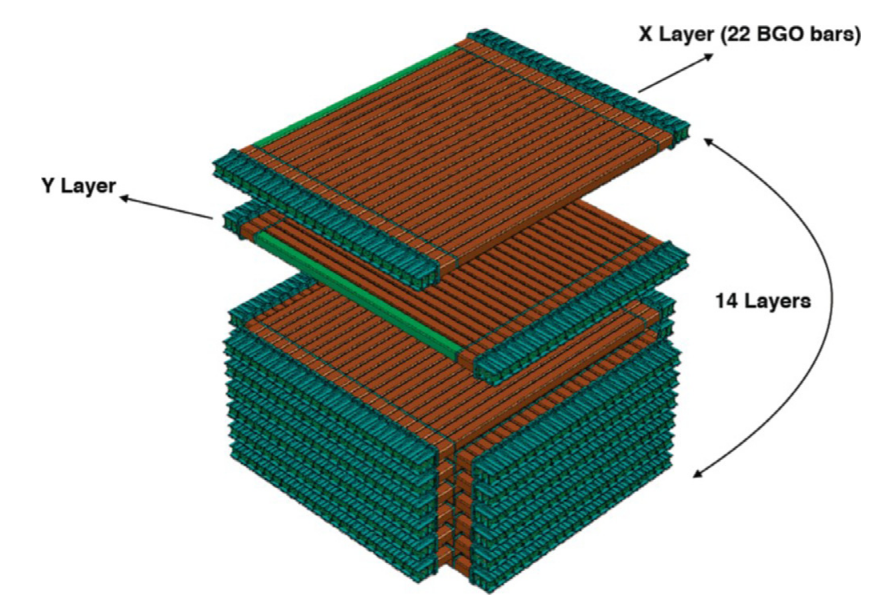}
  \caption{schematic view of the DAMPE calorimeter \cite{DAMPE_mission}.\label{fig:DAMPE_calorimeter}}
\end{figure}

Figure \ref{fig:DAMPE_calorimeter} shows a view of the DAMPE calorimeter, with its 14 layers and 22 BGO bars per layer \cite{DAMPE_mission, DAMPE_BGO}.
The bars have dimensions of 2.5$\times$2.5$\times$60 cm$^3$ and are alternately arranged along the $x$- and $y$- directions, resulting in a total active area on the vertical axis of 60$\times$60 cm$^2$.
The energy deposited in the bars is measured through scintillation by photomultiplier tubes (PMTs), coupled to both ends of a bar---namely S0 and S1---, using optical fibres.
The attenuation at the S1 end is five times higher than the one at S0, making the former more suitable to measure higher energies.
The PMTs used in DAMPE calorimeter are the R5610A-01 type produced by Hamamatsu, and they provide 10 different stages of charge amplification, or dynodes, identified as Dy1--10 \cite{DAMPE_PMTs}.
For the offline analysis, only signals from three of the available dynodes are recorded, specifically the ones from Dy2, Dy5 and Dy8.
The three channels have different gain factors, achieving a dynamic and wider range of measurable energies.
Dy8 has the highest gain, and therefore is the most sensitive to low signals; Dy2 has the lowest gain, and is the most appropriate to measure high energy deposits.
The Dy8 channels cover the range 2--500~MeV at the S0 end, and 10~MeV--2.5~GeV at S1; Dy5, 80~MeV--20~GeV at S0, and 400~MeV--100~GeV at S1; Dy2, 3.2~GeV--800~GeV at S0, and 16~GeV--4~TeV at S1 \cite{DAMPE_mission}.
When a signal read exceeds the maximum measurable value of the low-gain channel (i.e., Dy2) at the S1 end, the final reading is null.
In this situation, the bar is saturated and the acquired energy measurement is null.
For an example of a saturated event, see Figure \ref{fig:saturated_event}.

\begin{figure*}[htb]
  \centering
  \includegraphics[draft=false, width=0.9\textwidth]{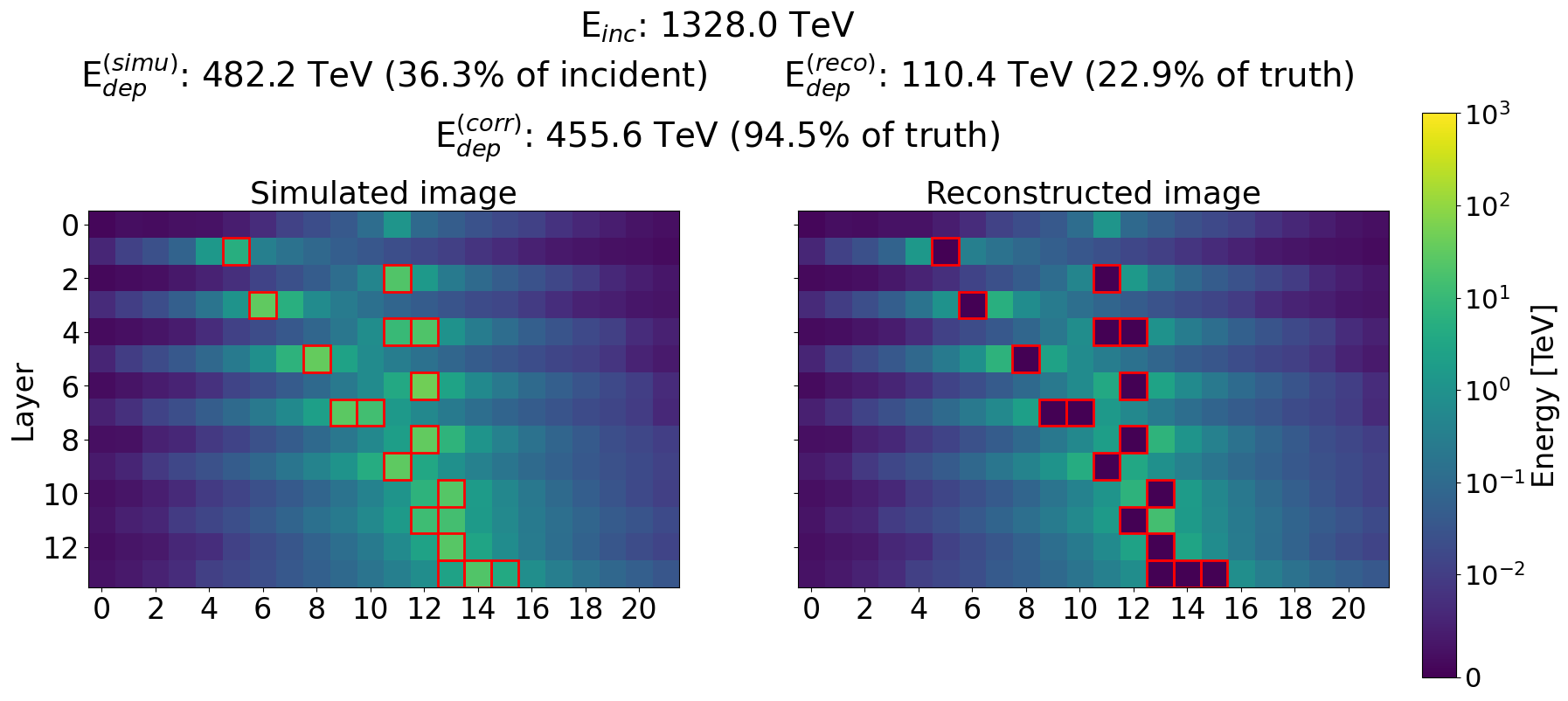}
  \caption{simulated event of an iron nucleus crossing the DAMPE calorimeter, with an incident energy of $\sim$1.3~PeV. The figure shows the simulated (on the left) and the reconstructed energy deposits (on the right) on the single BGO bars. Each subplot merges the information from both the layers with x- (odd layers) and y-oriented (even layers) bars. In the reconstructed event, null deposits resulting from saturation are present in the middle of the shower axis, where the release of energy is at its maximum.
  In both plots, the saturated bars are highlighted with red boxes. In the shown event, $\sim$80\% of the true deposited energy is lost because of saturation; after correction, more than 90\% of the true deposit is recovered.\label{fig:saturated_event}}
\end{figure*}

\section{Method}\label{sec:method}

The energy lost due to saturation in the calorimeter can be recovered using several methods and approaches.
Previous works showed the feasibility of performing such task using analytical and machine-learning methods \cite{DAMPE_saturation_analytical, DAMPE_saturation_ML_2022}.
Although their success in recovering the lost energy, those methods need specific corrections for different ions and start losing accuracy above $\sim$500~TeV.
In this work, a new ML model is used to correct the total deposited energy, independently from the type of ion crossing the detector, and maintaining the accuracy up to few PeVs of incident energy.
These two features are fundamental for the analyses of the individual CR fluxes---as well as the all-particle spectrum---, and the probing of the knee region, where a change of the spectral index is expected.
To train and validate the model, different ions crossing the full DAMPE experiment are simulated.

\subsection{Event simulation and selection}\label{sub:events}

Ions crossing the DAMPE instrument are simulated with the version 10.5 of the GEANT4 toolkit \cite{GEANT4}, using the \texttt{FTFP\_BERT} physics list \cite{GEANT4-PhysicsLists}; for primary energies above 100~TeV, the EPOS-LHC hadronic physics model is used, interfaced to GEANT4 through the Cosmic Ray Monte Carlo (CRMC) package \cite{CRMC,EPOS-LHC,DAMPE_simulations}.

Using the calorimeter hits, $\chi^2$ fits are independently performed in the $xz$- and $yz$- views to determine the particle trajectories.
To ensure a proper lateral containment in DAMPE calorimeter, and exclude non-reliable $\chi^2$ fits, a series of preliminary selections is applied to the simulated events; in particular:
\begin{itemize}
  \item no single layer should contain alone more than 35\% of the total energy;
  \item the maximum deposit for the three layers after the first one should not belong to the edge bars;
  \item the extrapolated positions of the reconstructed shower vector at the top and at the bottom of the calorimeter active volume should be within~280 mm from the volume center along the x- and y- directions.
\end{itemize}
These requirements are also in most cases applied to the on-orbit data to measure CR fluxes.
To improve the quality of the simulated samples, only events for which the true track is laterally contained in the active volume of the calorimeter are selected.
For the model training, events with a deposited energy inferior than 1~TeV are not considered; this condition is removed when studying the correction performances.

\subsection{Identification of events with saturation}\label{sub:identification_saturation}

An algorithm to identify saturated bars is developed to recognise events with saturation.
Simply considering bars with null deposits as saturated is not appropriate, because it does not discriminate the case where the ion does not cross the bar at all.
Saturation occurs in the middle of the shower profile, where the deposition is at its maximum.
According to this feature, the information of adjacent bars can be used to assess if an empty bar is accurate or saturated.
In this work, an empty bar is considered saturated when either its left or right neighbour on the same layer has a deposit greater than 25~GeV.
In a recurrent fashion, a bar is identified as saturated also when its deposit is null and a neighbour is saturated according to the previous condition; this is iterated until a non-null deposit or the edge of the layer is found.

\subsection{Model architecture}

Convolutional neural networks (CNNs) are typically used in ML applications for images processing \cite{CNN}.
A picture of width $\mathrm{W}$ and height $\mathrm{H}$ can be given as input to these models as an array of shape $\mathrm{W} \times \mathrm{H}$.
Similarly, the bar deposits of DAMPE calorimeter can be treated as a single-channel image of shape $14 \times 22$; the plots of Figure \ref{fig:saturated_event} give a visual representation of this.
Following this premise, a CNN can be designed to provide a correction factor for the total deposited energy, in the presence of saturation; this approach has already been proved successful in \cite{DAMPE_saturation_ML_2022}.
Figure \ref{fig:model_architecture} shows the architecture selected for this work.
The developed model can be divided into two main sections: first, a series of convolutional layers extracts features from the input image; then, the resulting output is flattened, and passed to a second series of dense layers, that provide the final correction factor in output.
The described ML architecture is implemented with TensorFlow 2.1.0 \cite{TensorFlow}.
A similar architecture is also used to track event candidates in DAMPE calorimeter, showing how this type of models can perform well for other regression tasks \cite{DAMPE_ML_tracking}.

\begin{figure*}[htb]
  \centering
  \includegraphics[draft=false, width=\linewidth]{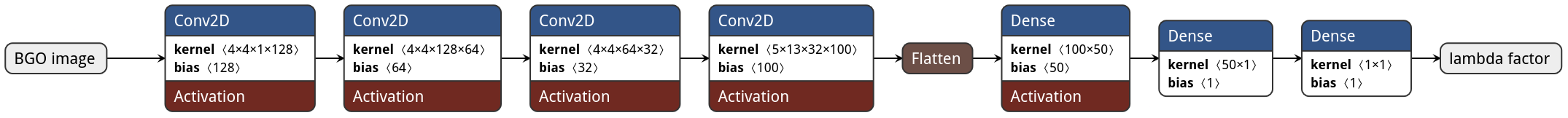}
  \caption{design of a convolutional neural network (CNN) to correct the total energy deposited in the DAMPE calorimeter for saturation. The specific model takes as input a $14 \times 22 \times 1$ image---where each pixel corresponds to a bar in the BGO calorimeter---, and returns a correction factor for the total energy.\label{fig:model_architecture}}
\end{figure*}

\subsection{Model training}

To generalise the correction for different ions, a model is trained using simulations of protons, helium, lithium, beryllium, boron, carbon, nitrogen, oxygen, neon, magnesium, silicon and iron nuclei.
Protons, helium, carbon, oxygen and silicon events range between 100~TeV and 1~PeV, iron between 100~TeV and 3~PeV, and the remaining elements between 100~TeV and 500~TeV.
The model is provided with all the events, without specifically selecting saturated events, so that it can also learn when no correction is needed.

ML models generally perform better when the given input values belong to the same range.
Several normalisations can be applied to the inputs for this purpose; for this work, the model is trained on BGO images divided by their maximum bar deposit.
In the same perspective, the CNN is trained to predict the target value
\begin{equation}\label{eq:lambda}
  \lambda = \mathrm{ln}\left(\frac{E_\mathrm{dep}^{(\mathrm{simu})}}{E_\mathrm{dep}^{(\mathrm{reco})}}\right)\text{,}
\end{equation}
with $E_\mathrm{dep}^{(\mathrm{simu})}$, $E_\mathrm{dep}^{(\mathrm{reco})}$ respectively the simulated and reconstructed total energies deposited in the calorimeter.

The full dataset contains a total of 2.13 million events, which are split into training (1.92 million events; i.e., 90\% of the full dataset) and validation sample (0.21 million events; i.e., 10\% of the full dataset).
The validation sample is needed to verify that the model is not overtrained.
If the overtraining occurs, the model loss on the validating events starts increasing, while the loss on the training sample keeps decreasing.
In that case, the model starts overfitting the training events and loses in generalisation when applied to a different set of events of the same kind.
Figure \ref{fig:training_loss} shows the model loss evaluated on both the training and validation samples, during training.
The loss is shown as a function of the training epochs, i.e., an iteration over the full set of training events.
The model training is stopped at the point where the loss function does not further decrease.
At most, 50 epochs are needed for the model to converge.

\begin{figure}[htb]
  \centering
  \includegraphics[width=\ifoptiondraft{0.5\linewidth}{0.8\linewidth}, draft=false]{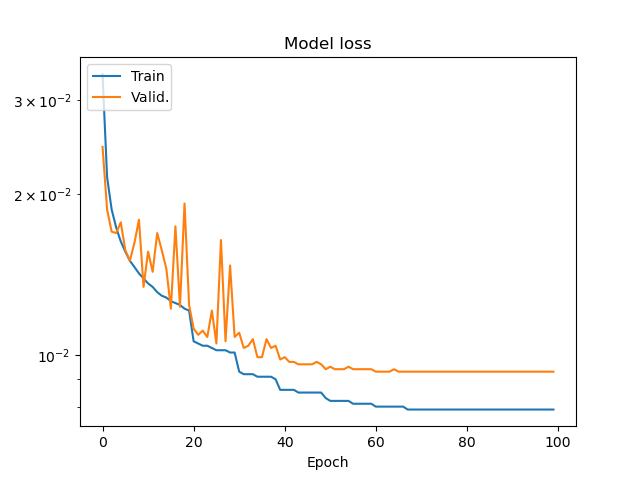}
  \caption{loss of the model during training. The loss function is evaluated at the end of each epoch on both the training and the validation samples. The decreasing trend found in both cases excludes the presence of overtraining.\label{fig:training_loss}}
\end{figure}

\section{Results}

Simulations of protons, carbon and iron nuclei with primary energies between 100~TeV--3~PeV crossing the DAMPE calorimeter are used to test the behavior of the trained model.
Figure \ref{fig:sat_frac} shows the fraction of saturated events present in the considered samples.
Figures \ref{fig:prim}--\ref{fig:nsat} show the ratio of the total reconstructed energy after correction and the corresponding true simulated value, as a function of the incident energy, the fraction of energy lost in saturation and the number of identified saturated bars.
With reference to the Figures \ref{fig:prim} and \ref{fig:nsat}, the primary energy and the number of saturated bars do not substantially affect the accuracy of the correction.
On the contrary, Figure \ref{fig:miss} shows how the fraction of missing energy impacts the predictions.
In particular, highly saturated events, with more than $\sim$90\% of the true deposit missing, show a stronger bias of $\sim-$20--30\%.
In other cases, the remainder of missing energy does not exceed 10\% of the true value.

\begin{figure*}[htb]
  \centering
  \includegraphics[draft=false, width=0.48\textwidth]{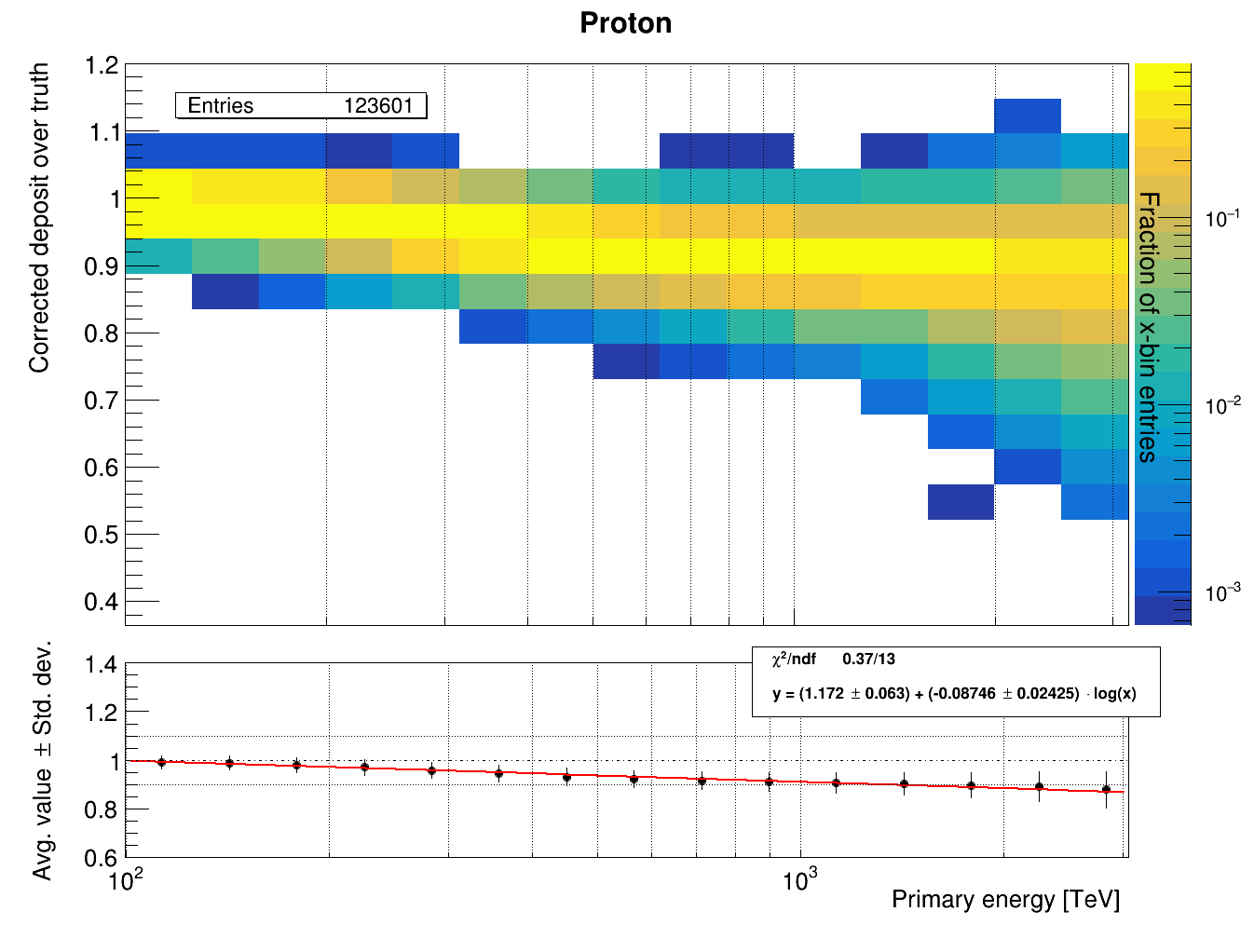}
  \includegraphics[draft=false, width=0.48\textwidth]{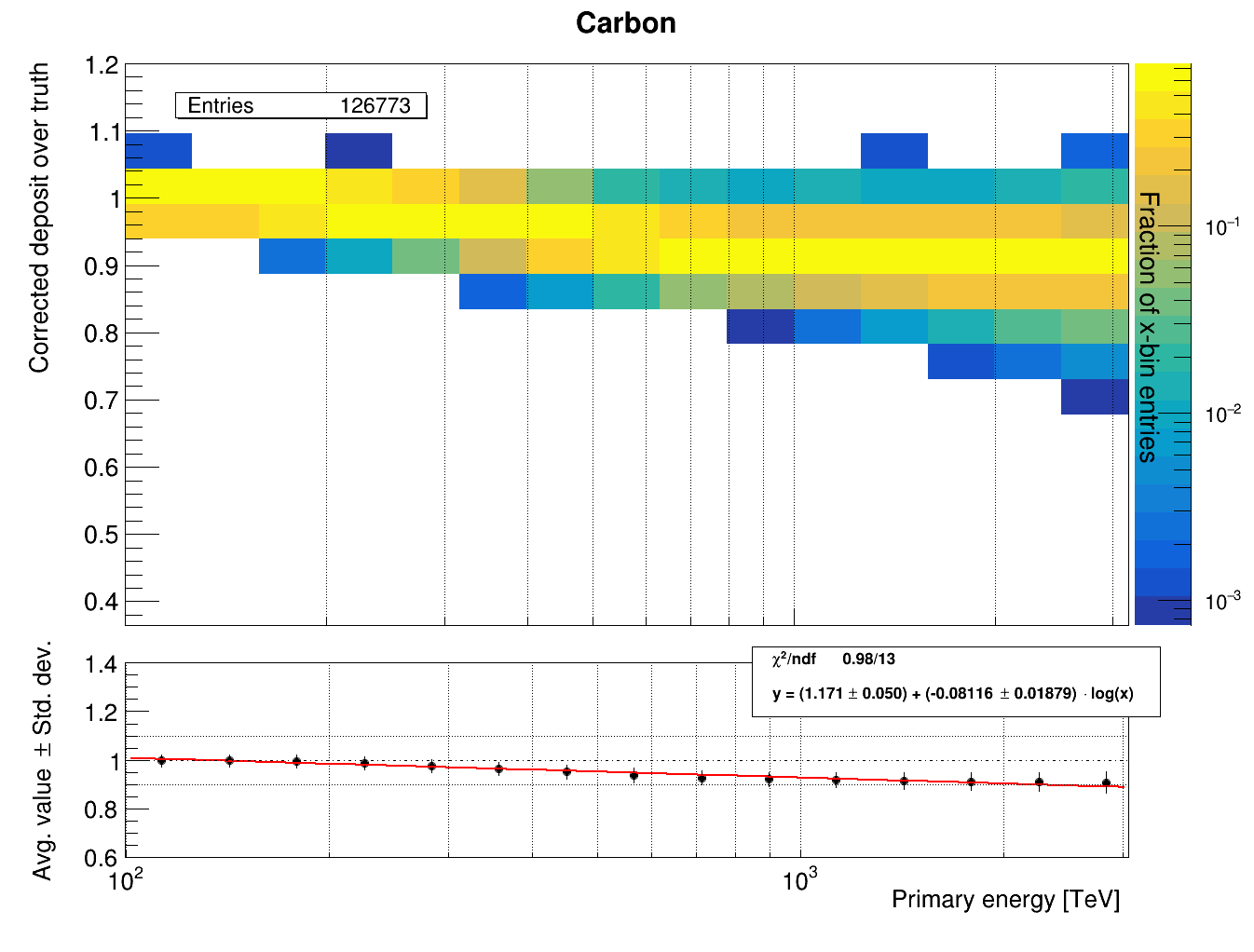}\\[3pt]
  \includegraphics[draft=false, width=0.48\textwidth]{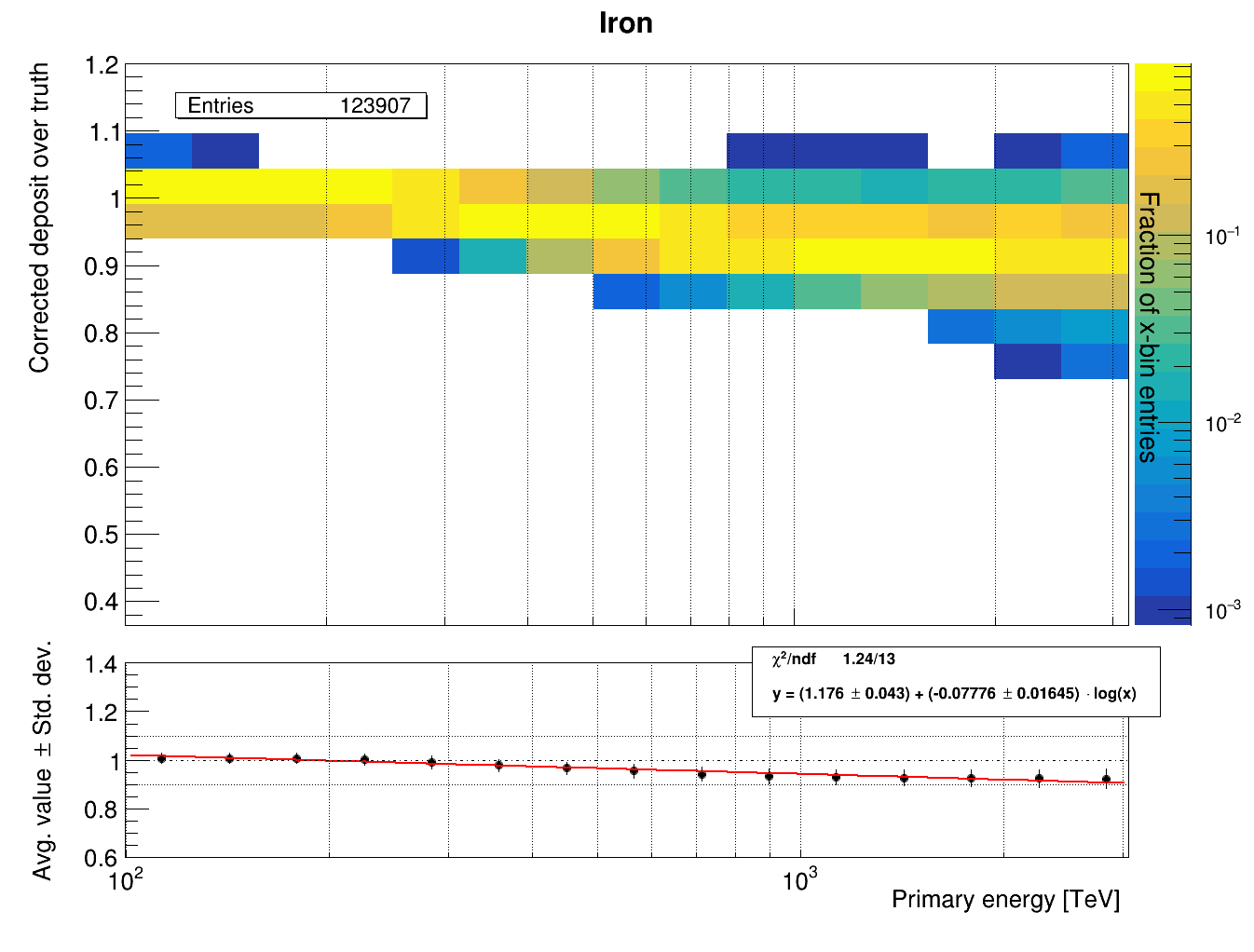}
  \caption{ratio of the corrected and simulated total deposited energy, as a function of the particle incident energy. Three plots are shown for the protons (top left), carbon (top right) and iron events (bottom); each plot is split into two subplots: the main one shows the 2D distribution of the events with saturation, and the secondary one the average value and the standard deviation of the y-projection of the corresponding x-bin, from the main plot. The values in the secondary subplot are fitted using a linear function to the 10-logarithm of the incident energy.\label{fig:prim}}
\end{figure*}
\begin{figure*}[htb]
  \centering
  \includegraphics[draft=false, width=0.48\textwidth]{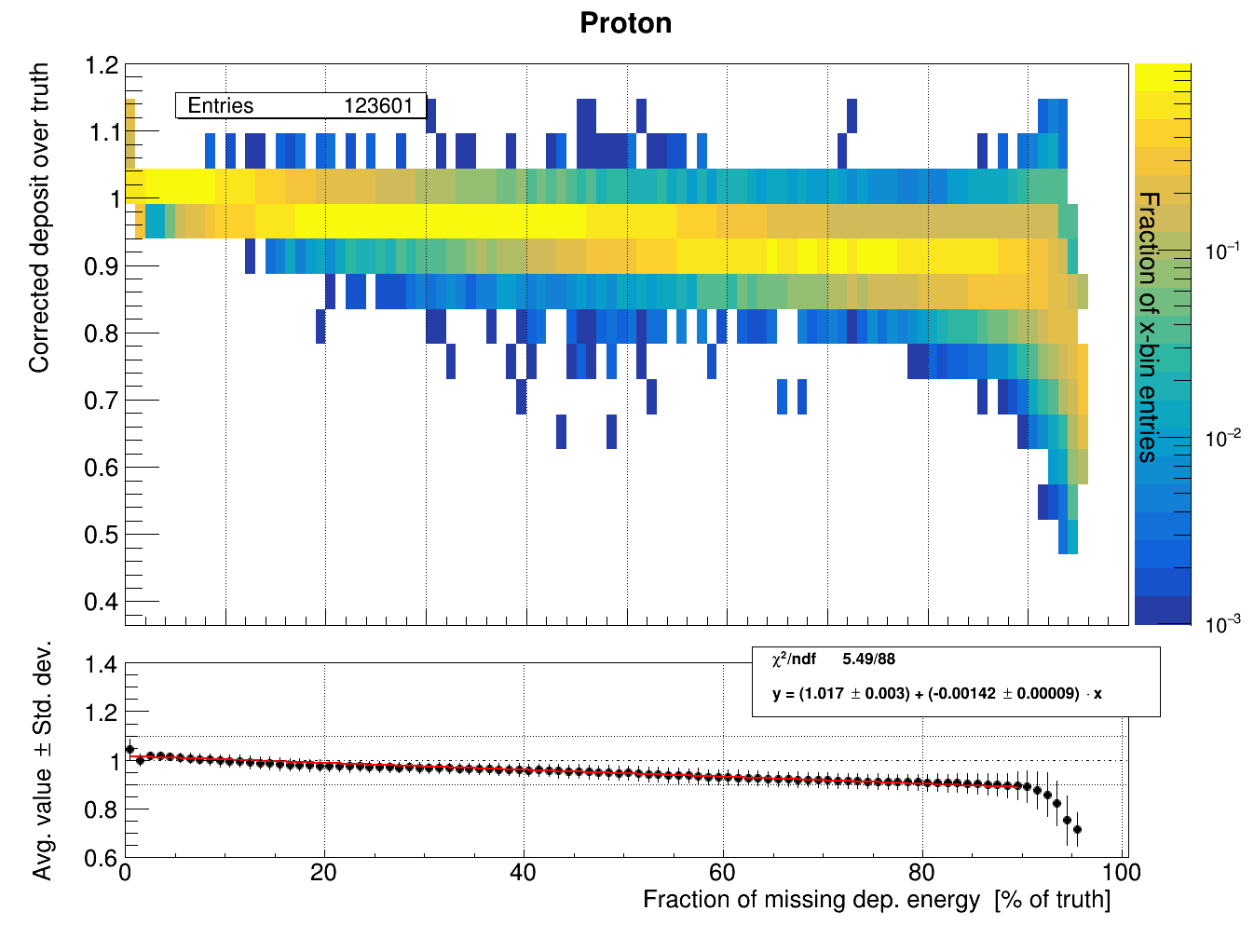}
  \includegraphics[draft=false, width=0.48\textwidth]{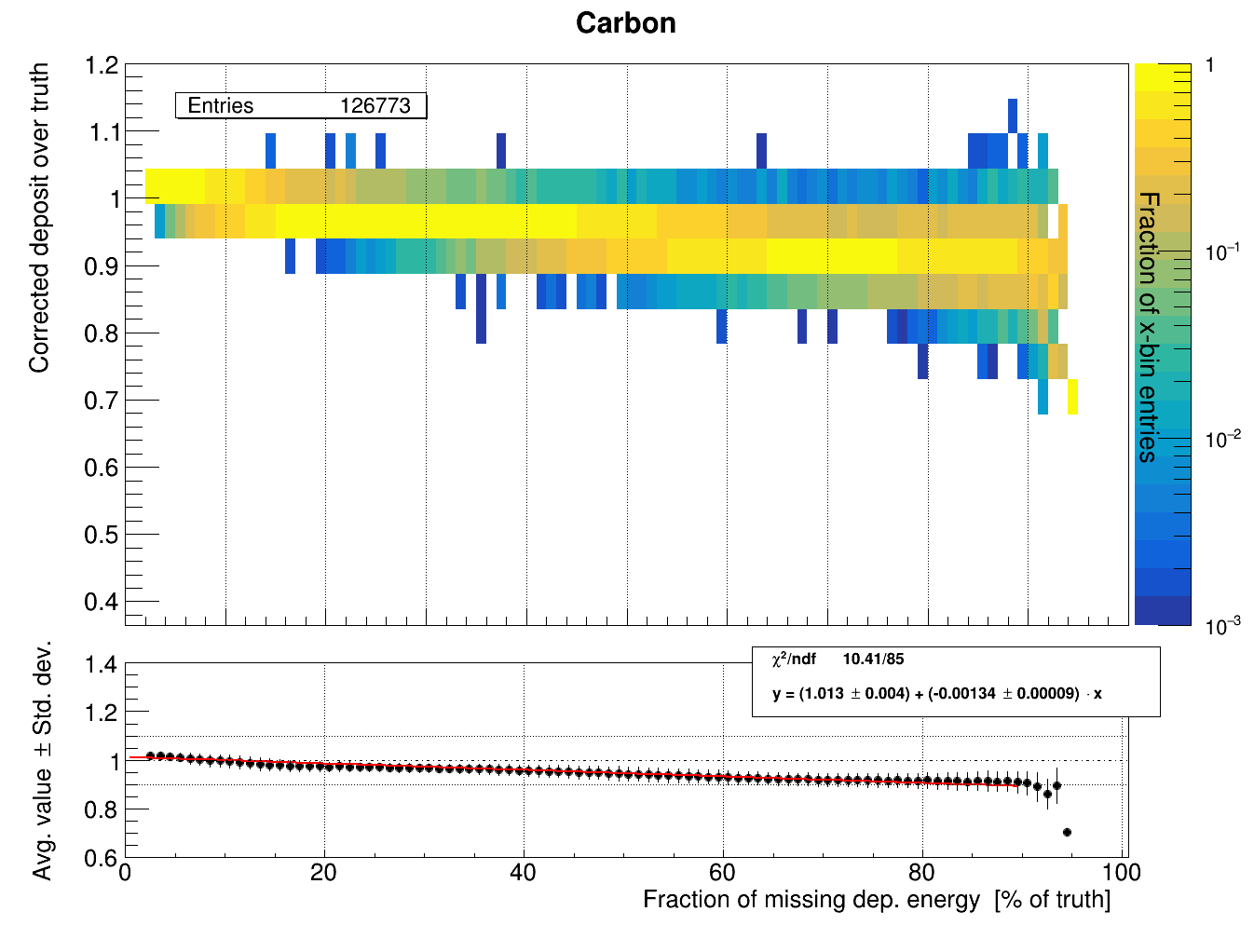}\\[3pt]
  \includegraphics[draft=false, width=0.48\textwidth]{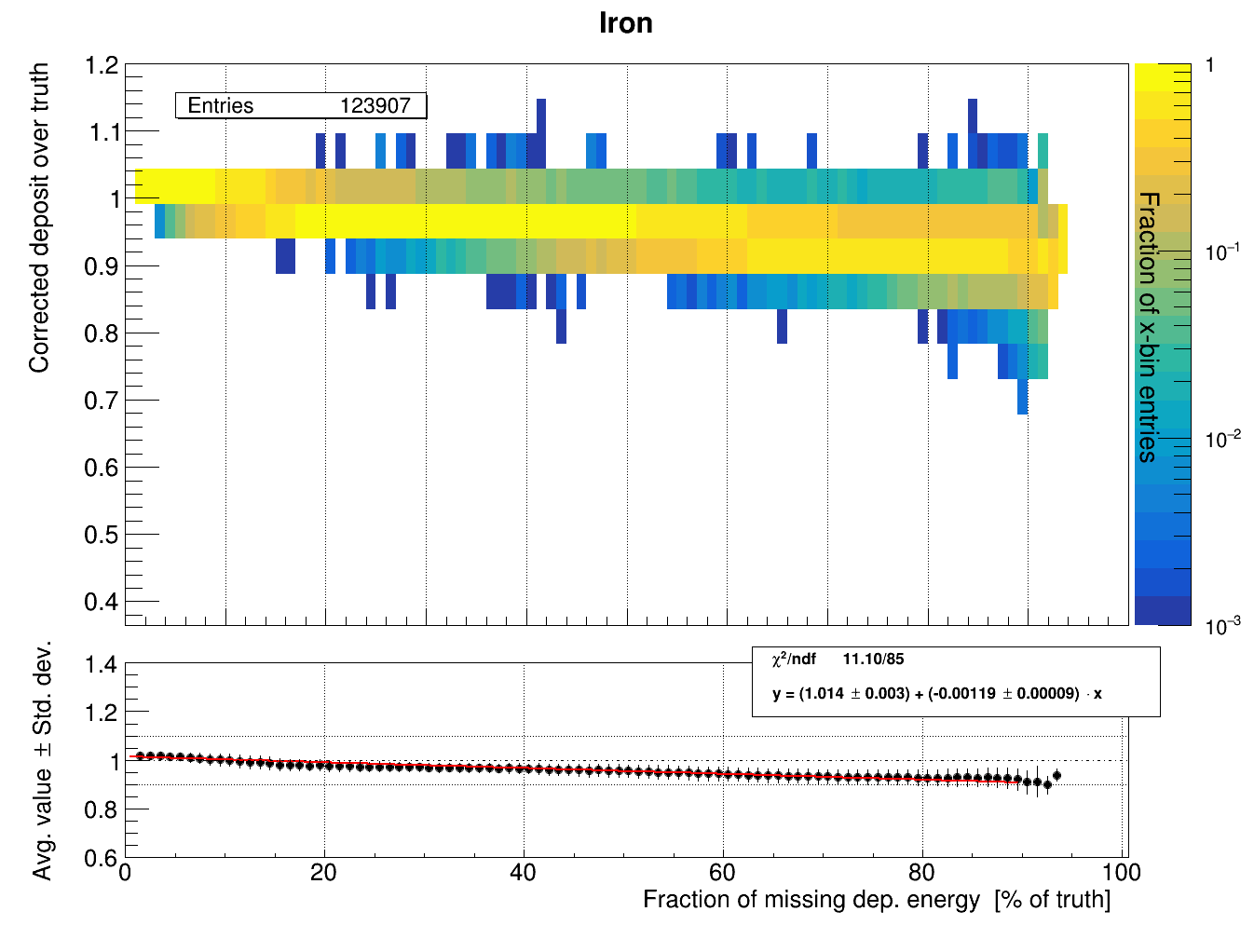}
  \caption{as per Figure \ref{fig:prim}, but as a function of the fraction of true energy lost in saturation. Differently from Figure \ref{fig:prim}, the values in the secondary subplots are fitted with a linear function to the fraction of missing energy.\label{fig:miss}}
\end{figure*}
\begin{figure*}[htb]
  \centering
  \includegraphics[draft=false, width=0.48\textwidth]{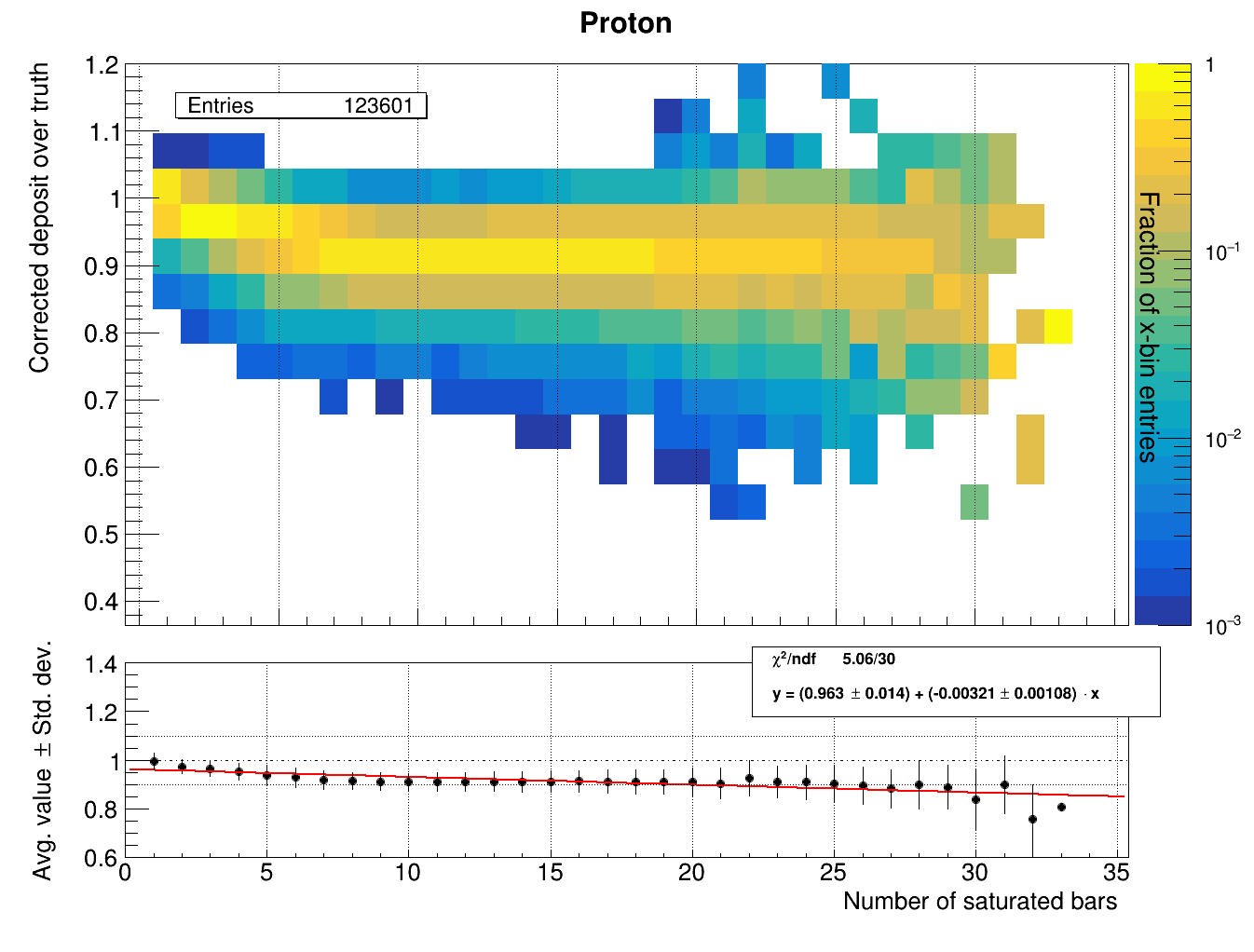}
  \includegraphics[draft=false, width=0.48\textwidth]{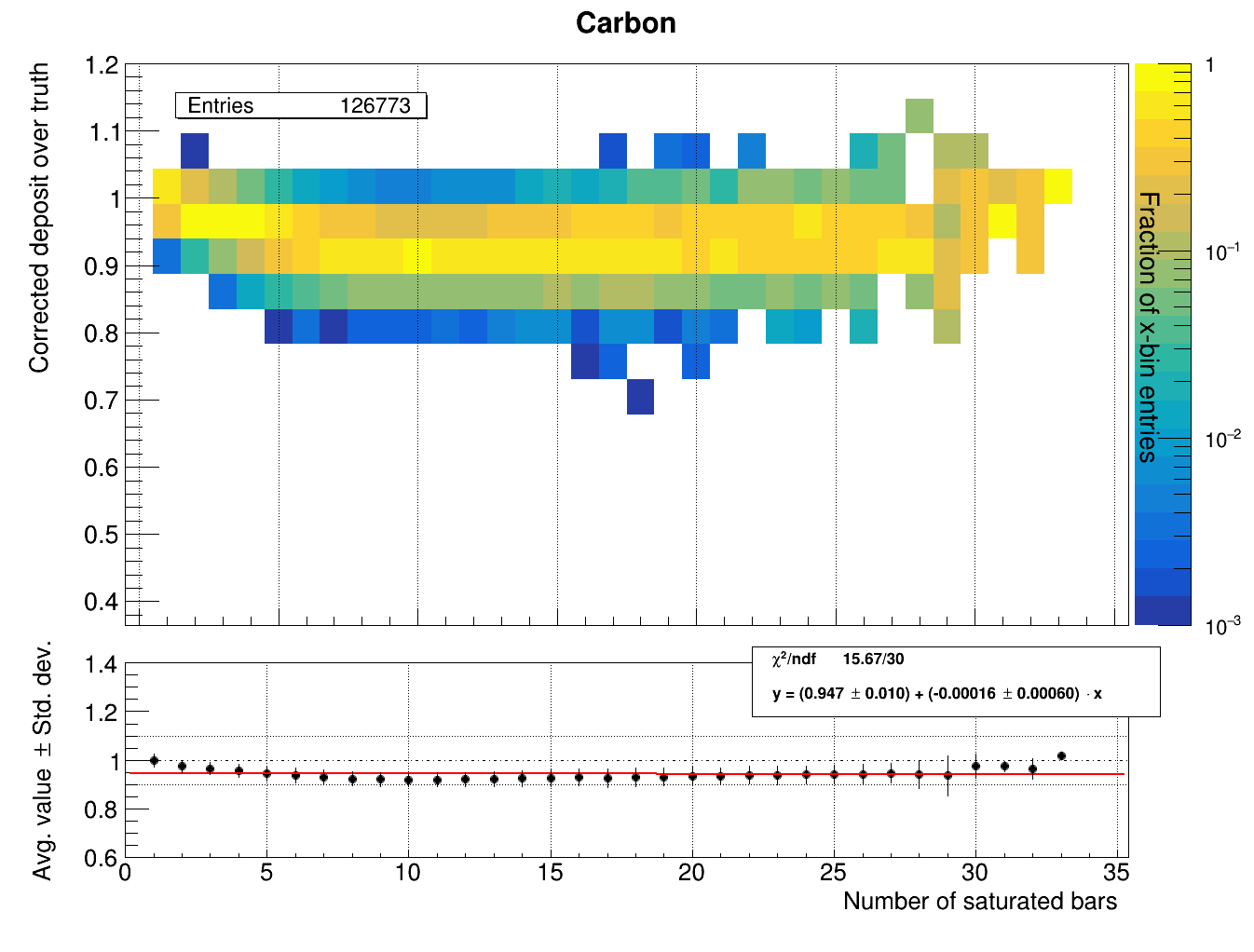}\\[3pt]
  \includegraphics[draft=false, width=0.48\textwidth]{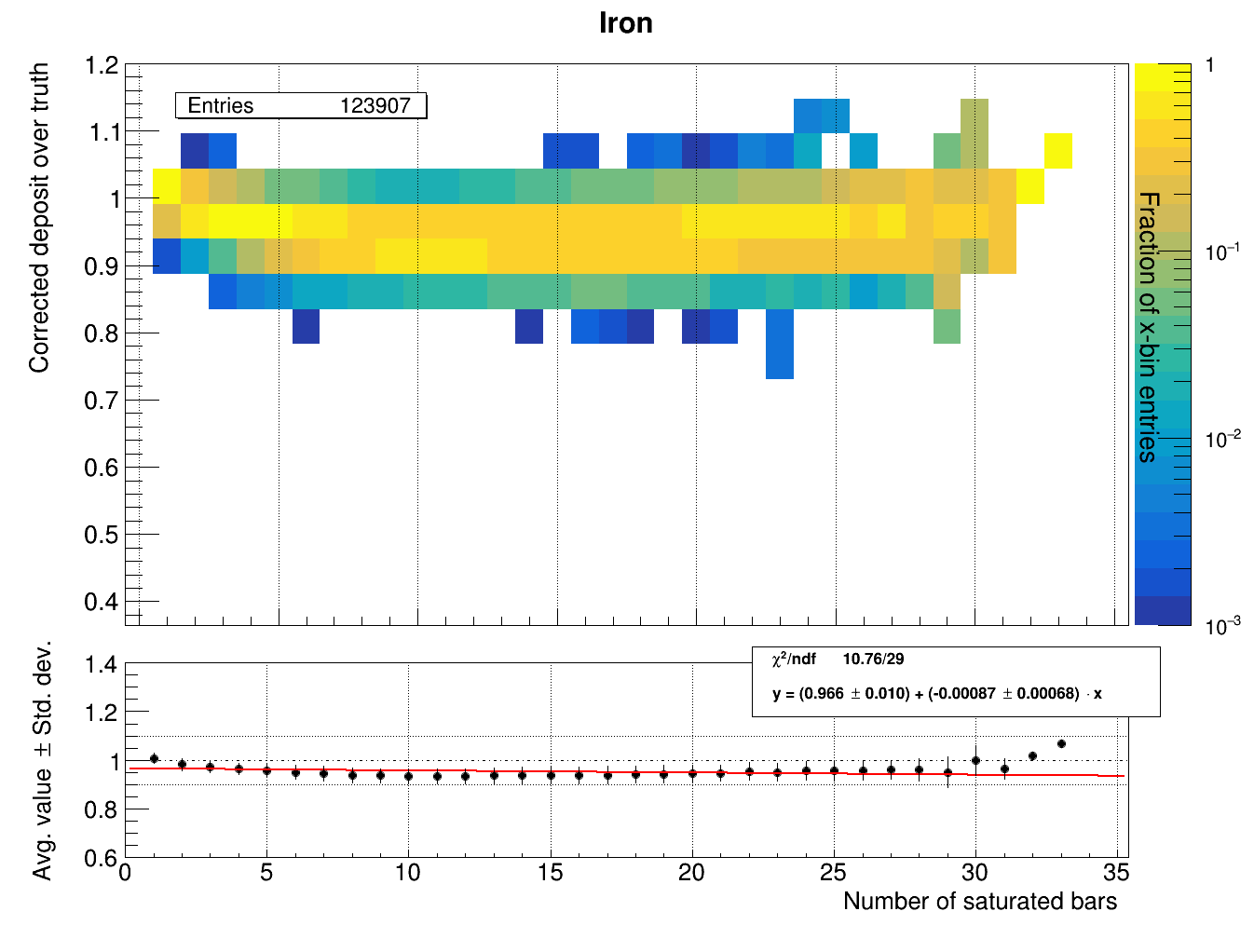}
  \caption{as per Figure \ref{fig:prim}, but as a function of the number of identified saturated bars. Differently from Figure \ref{fig:prim}, the values in the secondary subplots are fitted with a linear function to the number of saturated bars.\label{fig:nsat}}
\end{figure*}

In Figure \ref{fig:reweight} the events are re-weighted to a $E_\mathrm{inc}^{-2.7}$ spectrum, similarly to an actual CR flux, ignoring possible breaks in the spectral index \cite{PDG2024} (see \ref{app:events_reweighting} for further details).
The distribution of the total deposited energy before and after the correction shows that the residual bias for highly saturated events, seen in Figure \ref{fig:miss}, affects the very last energy bins.

\begin{figure*}[htb]
  \centering
  \includegraphics[draft=false, width=0.48\textwidth]{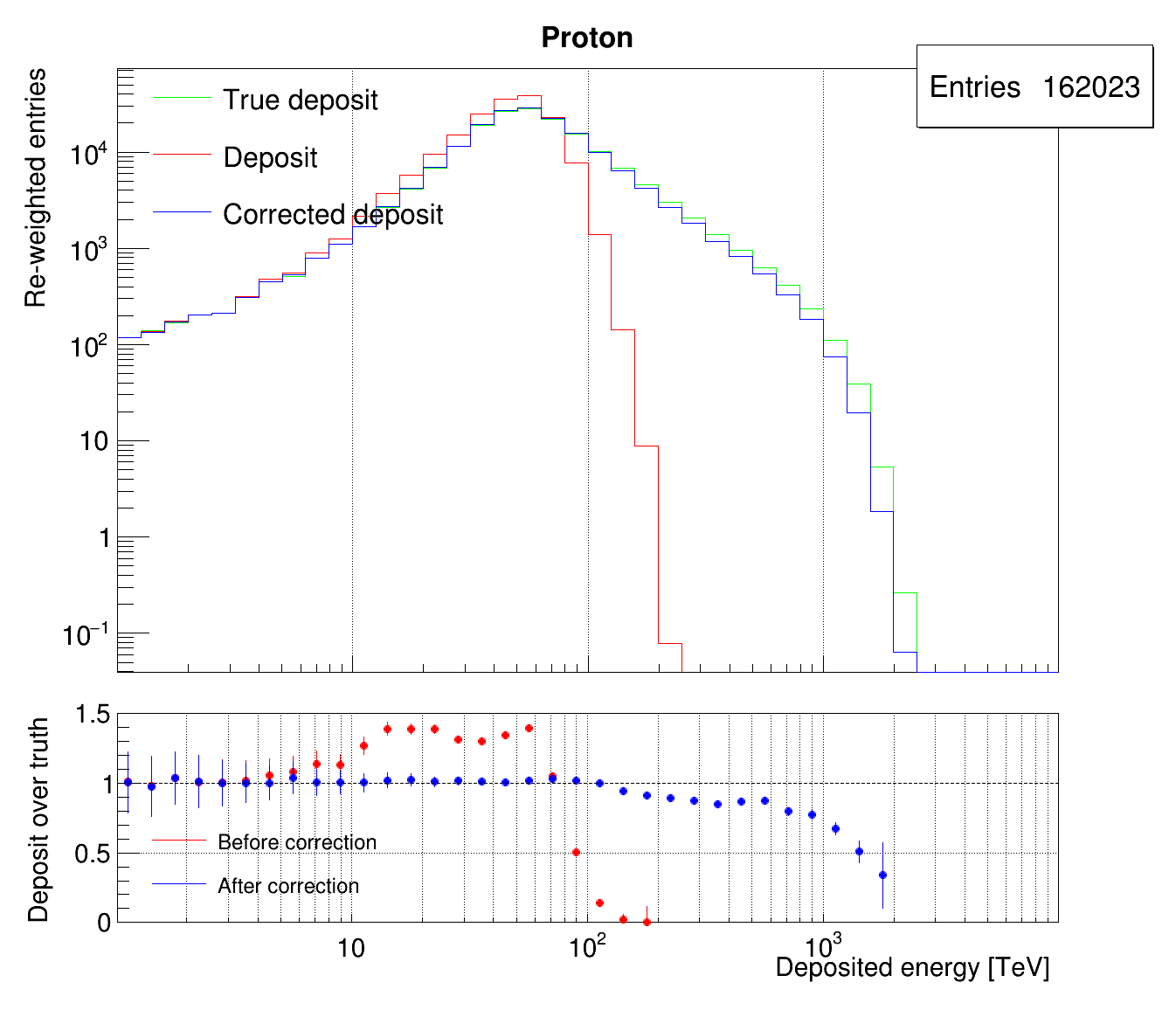}
  \includegraphics[draft=false, width=0.48\textwidth]{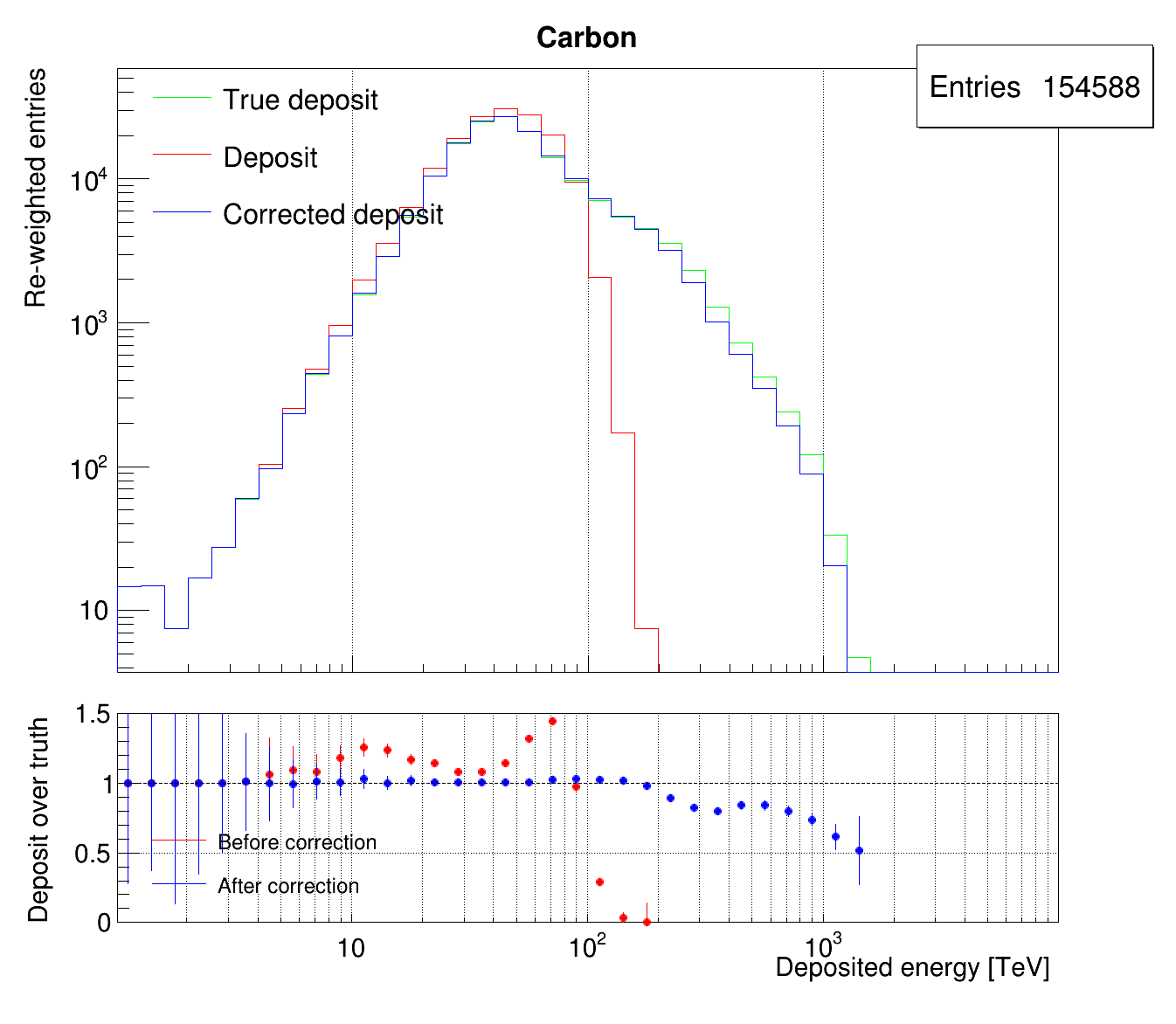}\\[3pt]
  \includegraphics[draft=false, width=0.48\textwidth]{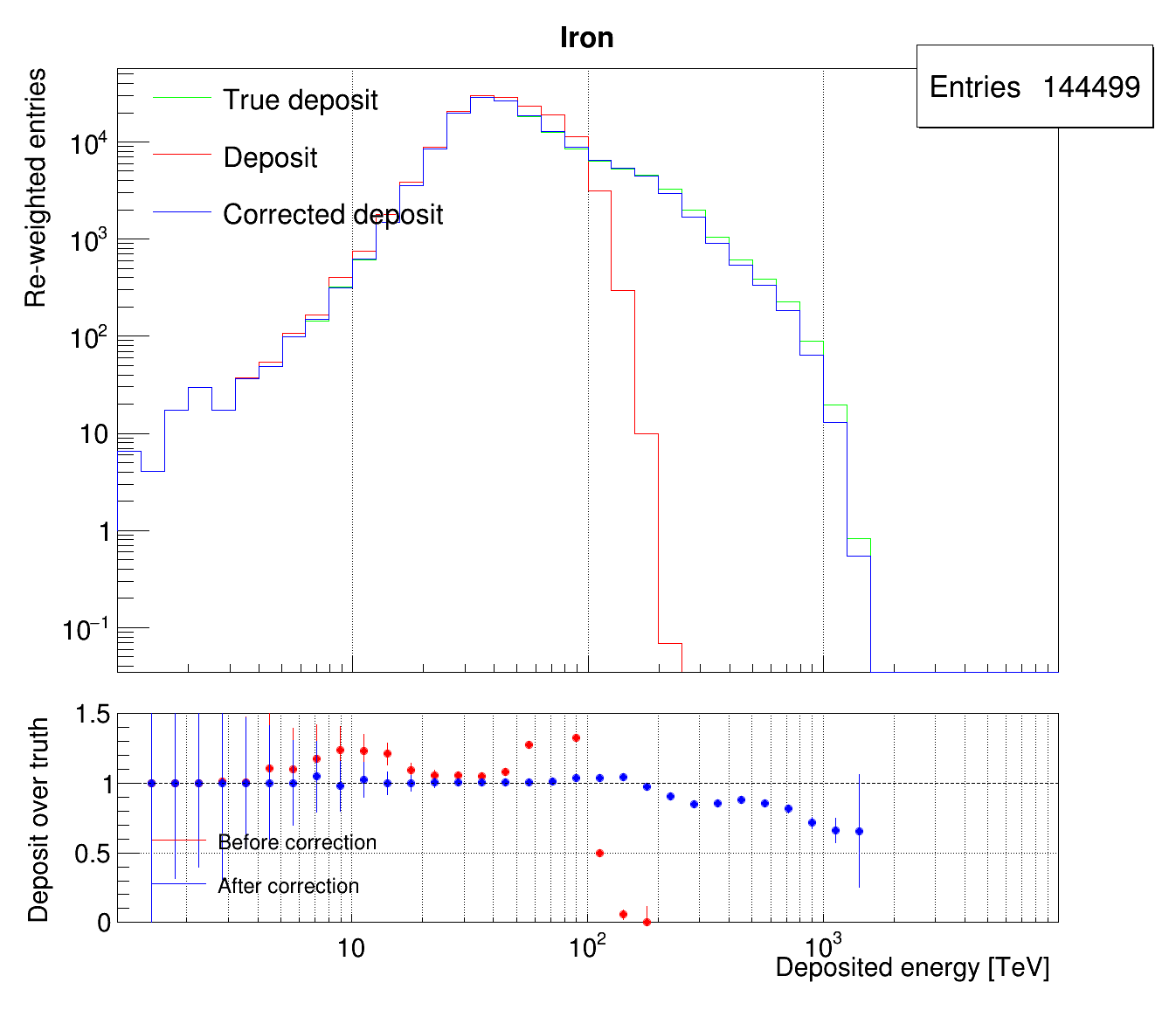}
  \caption{distribution of the total deposited energy of simulated events. The event counts are re-weighted to a $E_\mathrm{inc}^{-2.7}$ spectrum, based on their incident energy. Three plots are shown for the protons (top left), carbon (top right) and iron events (bottom); each plot is split into two subplots: the top one shows the distributions of the true value of the total deposited energy (green), and the reconstructed total deposit before (red) and after (blue) correction; the bottom one shows the ratio between the bin counts of the reconstructed and simulated energy, before (red) and after (blue) correction.\label{fig:reweight}}
\end{figure*}

For the other species used for training---within their trained energy range---, the model performance is consistent with these results.

\subsection{Testing with on-orbit data}

Despite the efforts, the physics in MC simulations might differ from the one in place during real data taking.
To proof the saturation correction, the behavior observed on MC events needs to be confirmed by comparison with real data.
Unlike simulations, data do not provide primary information about the lost energy, thus no ground truth is available.
However, it is possible to take non-saturated events and artificially saturate the bars with a deposit greater than a lowered saturation threshold.
For this work, a saturation minimum level is estimated for each bar from MC events, and bar deposits found in data to be greater than 80\% of the corresponding threshold are removed.
For the generated pseudo-saturated events, the saturation correction is applied to the manipulated image, and the corrected total deposit is compared to the originally measured value, which is now treated as the truth.

Pseudo-saturated data entries are selected from the on-orbit events acquired between 2016 and 2024, with a measured deposited energy greater than 1~TeV.
The PSD is used to measure the charge of the traversing particles and identify different ions in data.
In particular, protons are selected with an average charge in between 0.8 and 1.3, carbon nuclei in between 5.8 and 6.3, and iron nuclei in between 25.8 and 26.3.
For more information on how the average charge is measure, refer to \ref{app:psd_charge}.
Proton, carbon and iron pseudo-saturated data samples cover the range of 10--100~TeV with their originally measured deposited energy (see Figure \ref{fig:pseudo_simu}).
To cross-check the results obtained on simulations with data, also simulated events are pseudo-saturated.
Since the primary information is not available in data, the lateral containment requirement on the true track is lifted from both MC and data samples.

Figures \ref{fig:pseudo_simu}--\ref{fig:pseudo_nsat} show the ratio of the corrected total energy and the corresponding true value for MC and data pseudo-saturated events, as a function of the total measured deposited energy, the fraction of energy lost in pseudo-saturation and the number of artificially saturated bars.
In this case, the known number of artificially saturated bars is used, instead of the counts obtained with the algorithm described in subsection \ref{sub:identification_saturation}.

\begin{figure*}[htb]
  \centering
  \includegraphics[draft=false, width=0.48\textwidth]{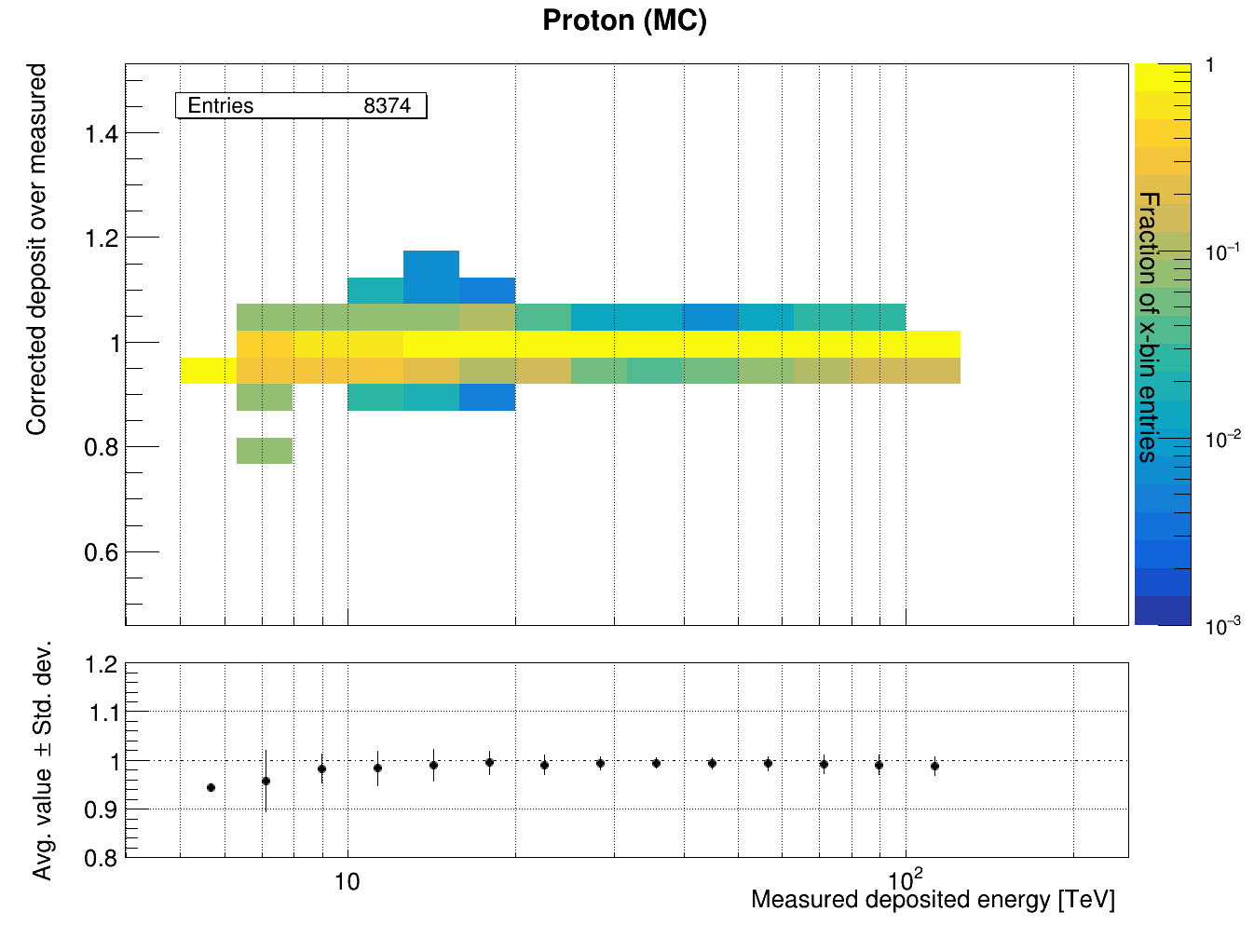}
  \includegraphics[draft=false, width=0.48\textwidth]{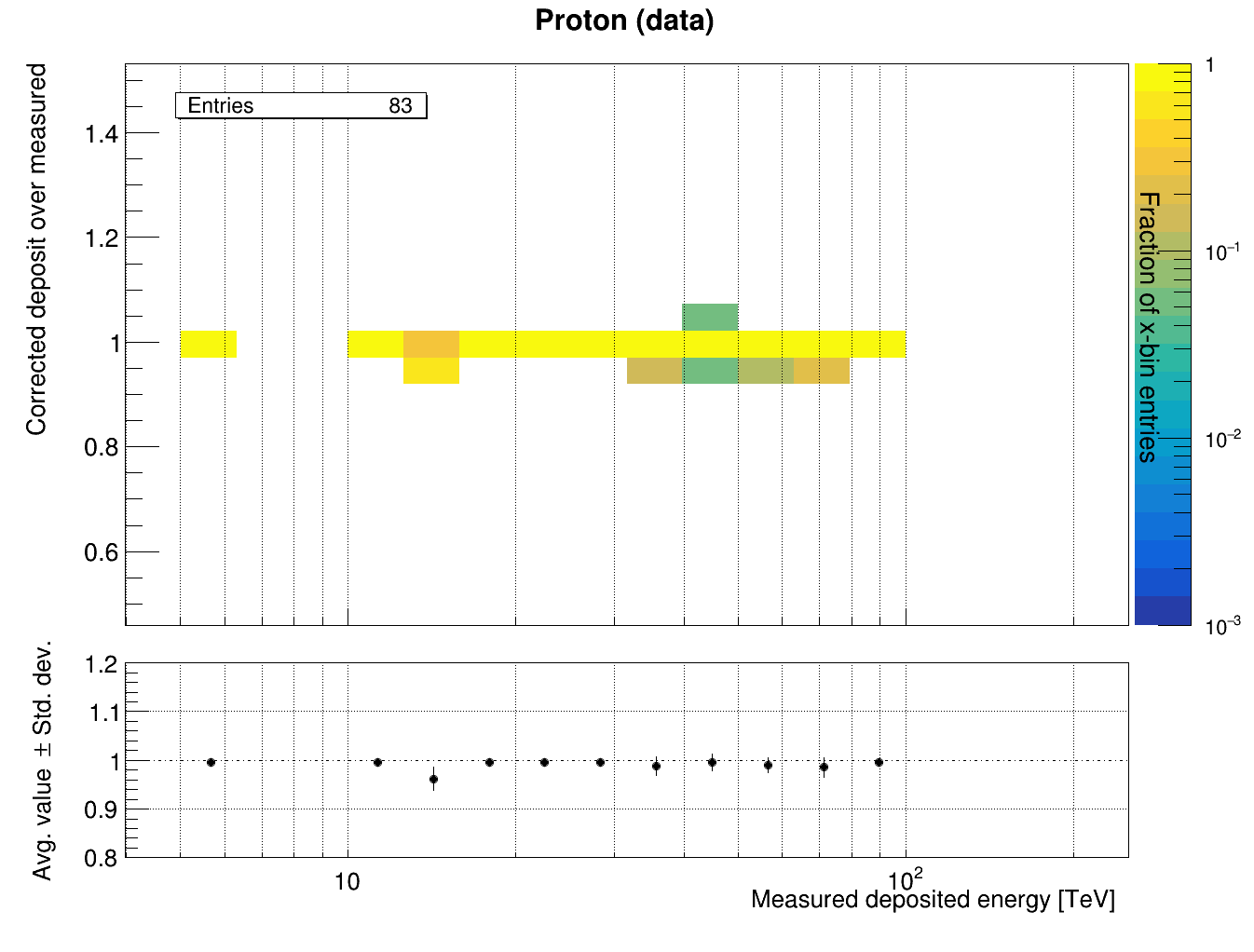}\\[3pt]
  \includegraphics[draft=false, width=0.48\textwidth]{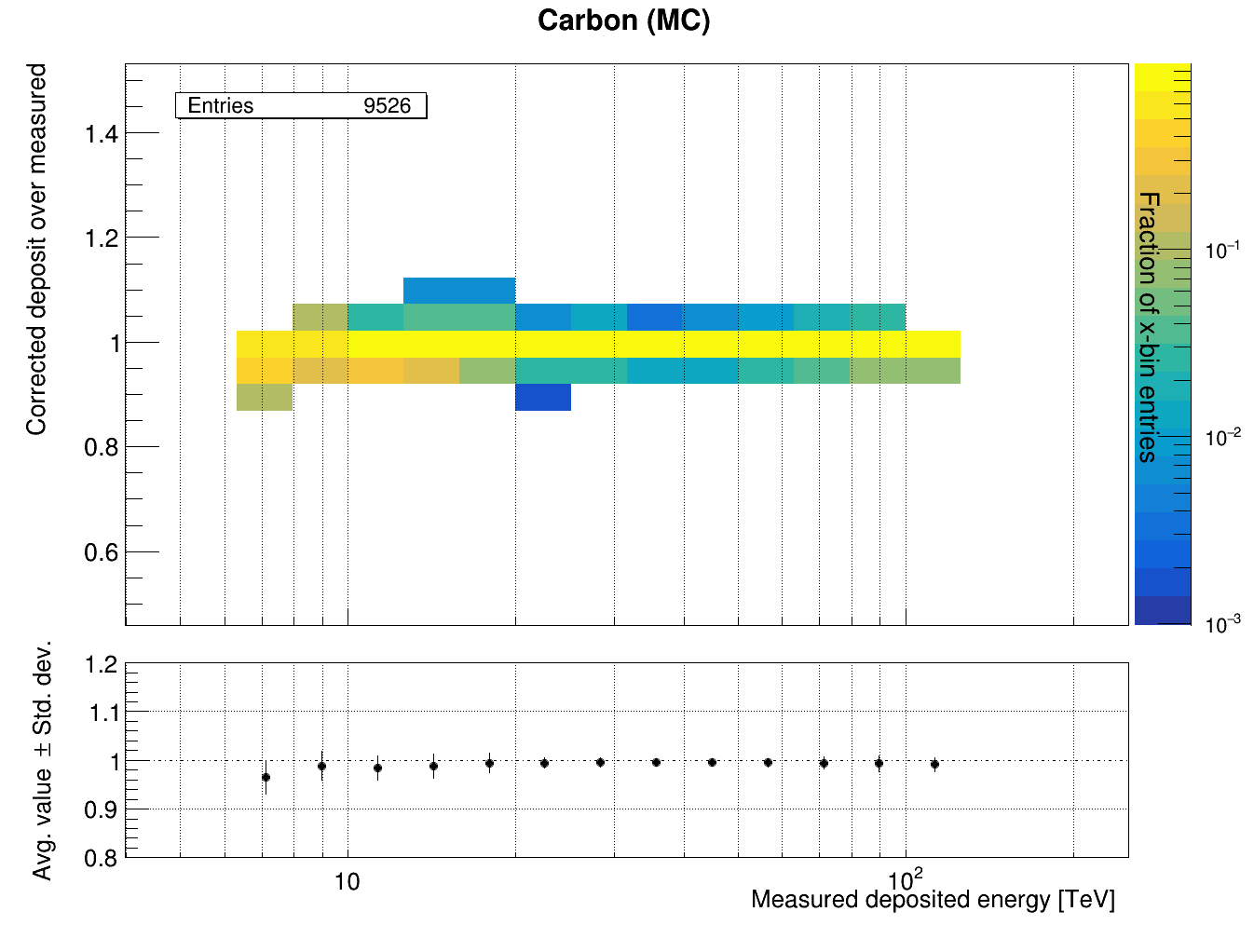}
  \includegraphics[draft=false, width=0.48\textwidth]{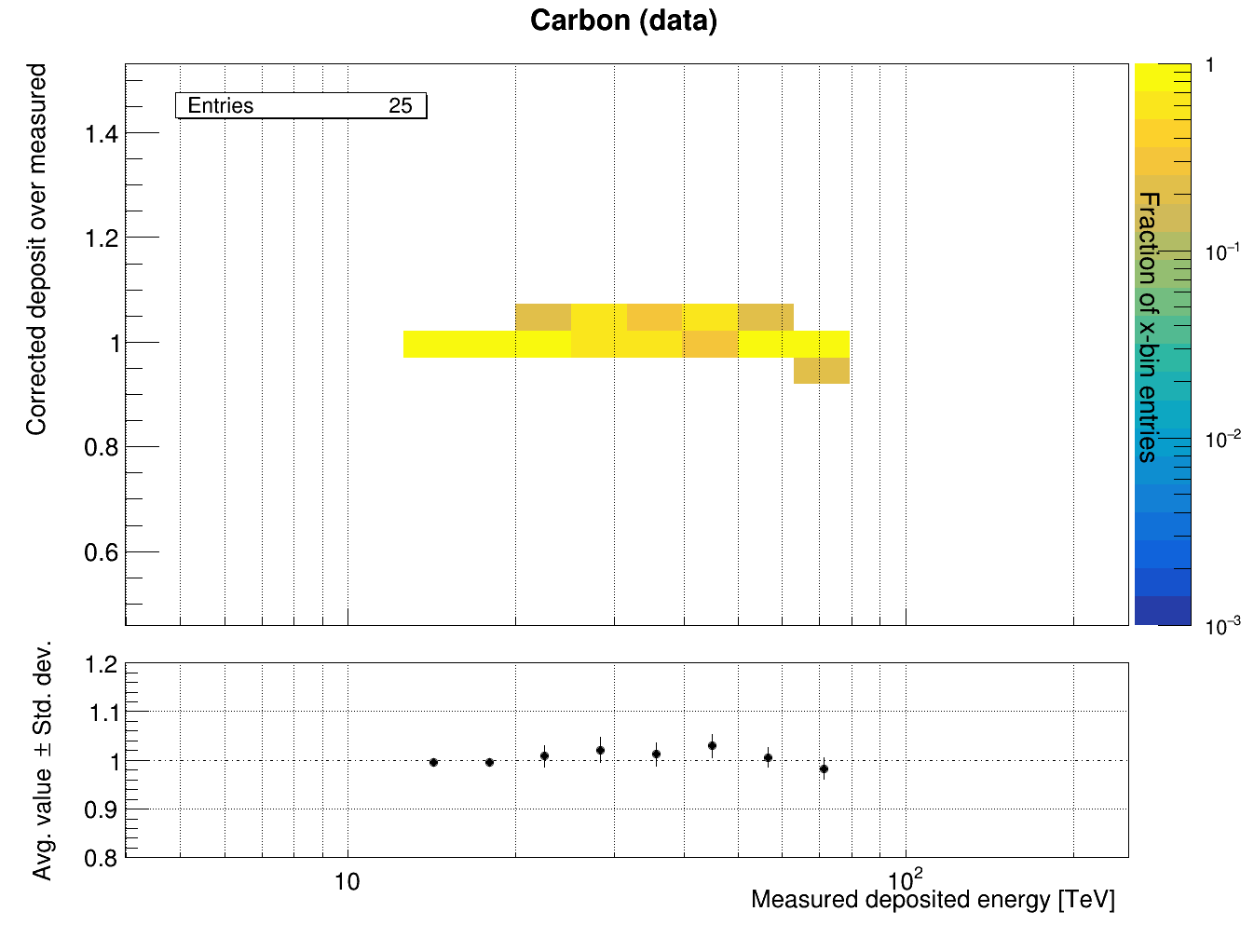}\\[3pt]
  \includegraphics[draft=false, width=0.48\textwidth]{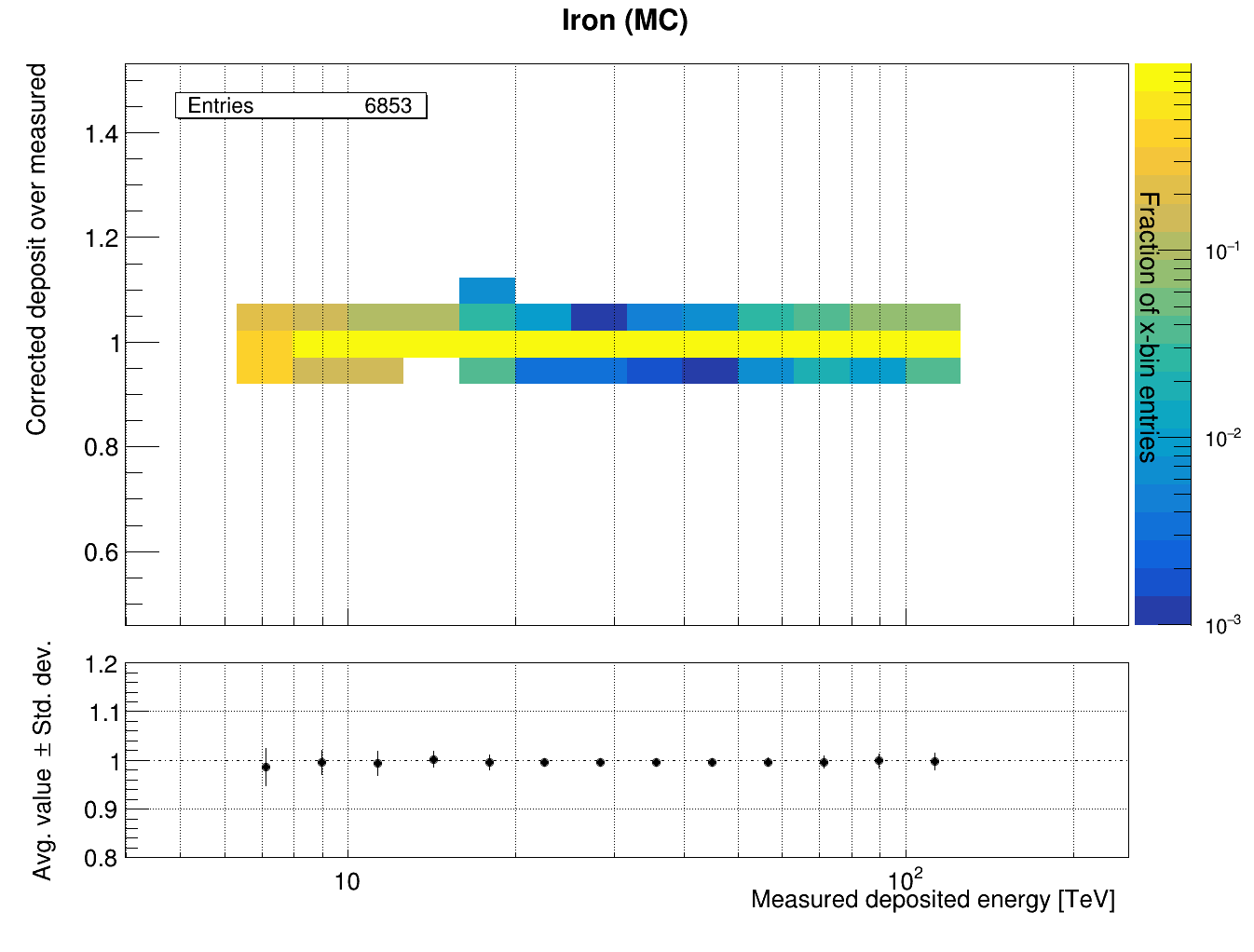}
  \includegraphics[draft=false, width=0.48\textwidth]{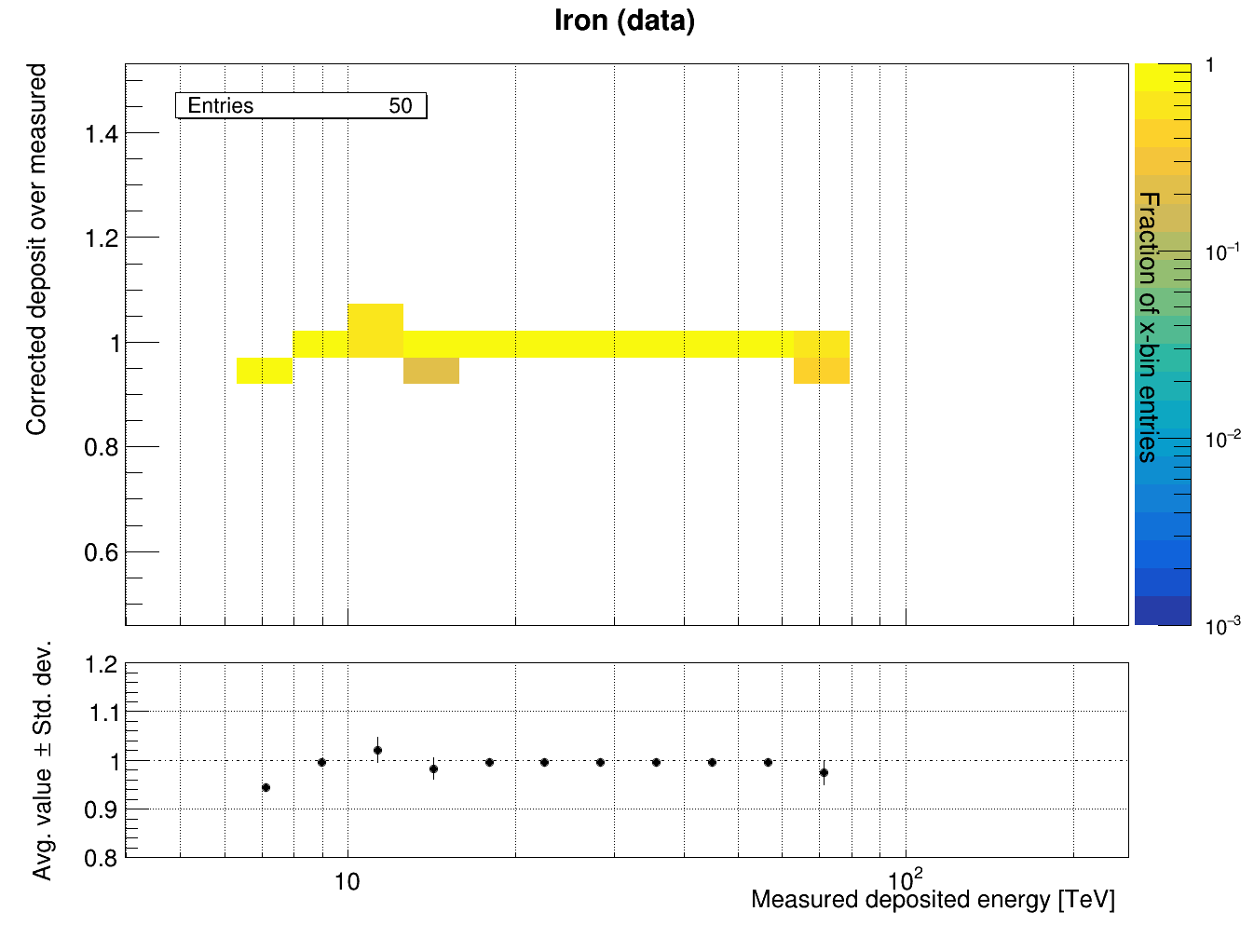}
  \caption{ratio of the corrected and the total measured deposited energy for pseudo-saturated events, as a function of the total measured deposited energy. The left column shows MC events, while the right column real data; from top to bottom, proton, carbon, and iron nuclei are selected. Each plot is split into two subplots: the main one shows the events 2D distribution, and the secondary one the average value and the standard deviation of the y-projection of the corresponding x-bin, from the main plot.\label{fig:pseudo_simu}}
\end{figure*}
\begin{figure*}[htb]
  \centering
  \includegraphics[draft=false, width=0.48\textwidth]{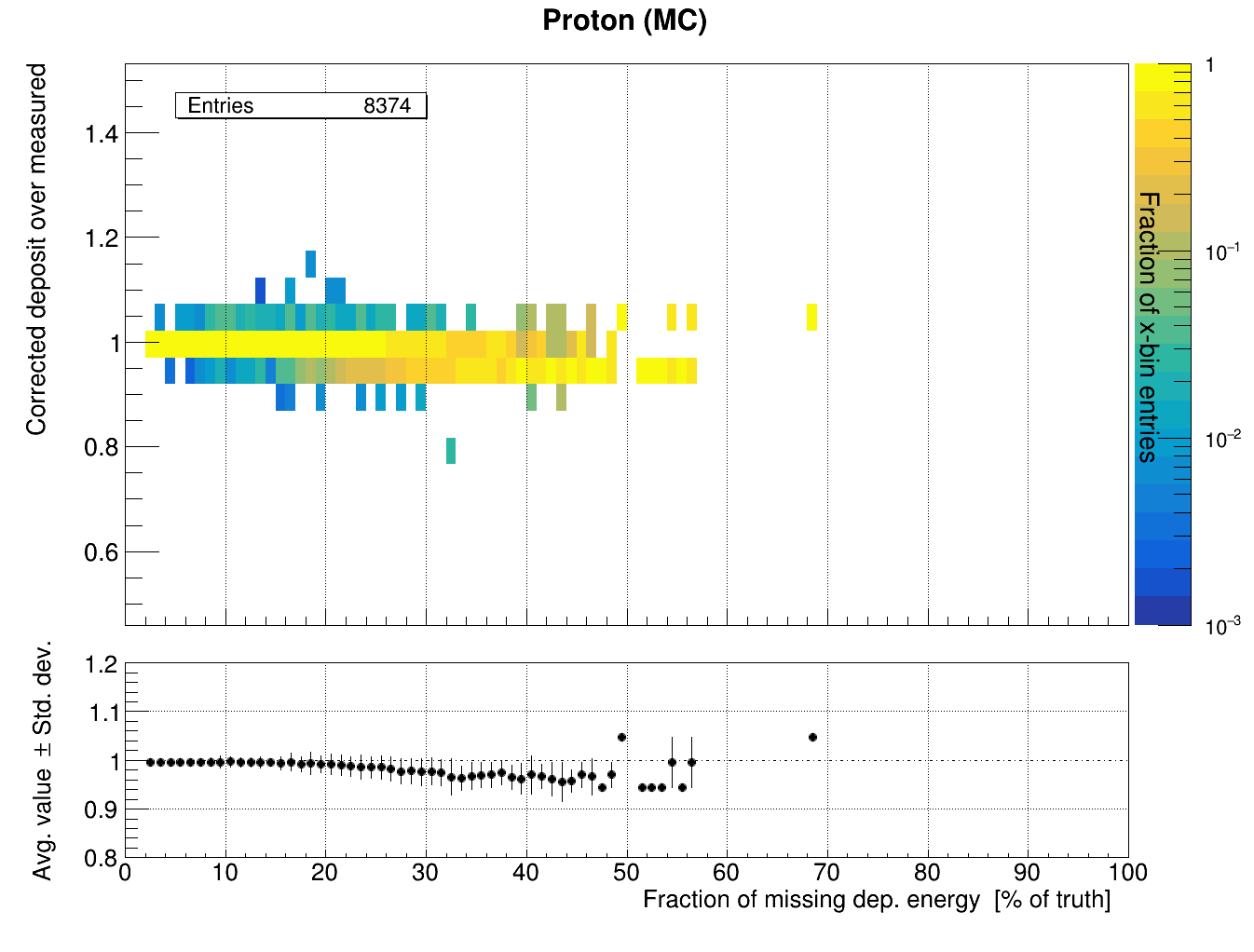}
  \includegraphics[draft=false, width=0.48\textwidth]{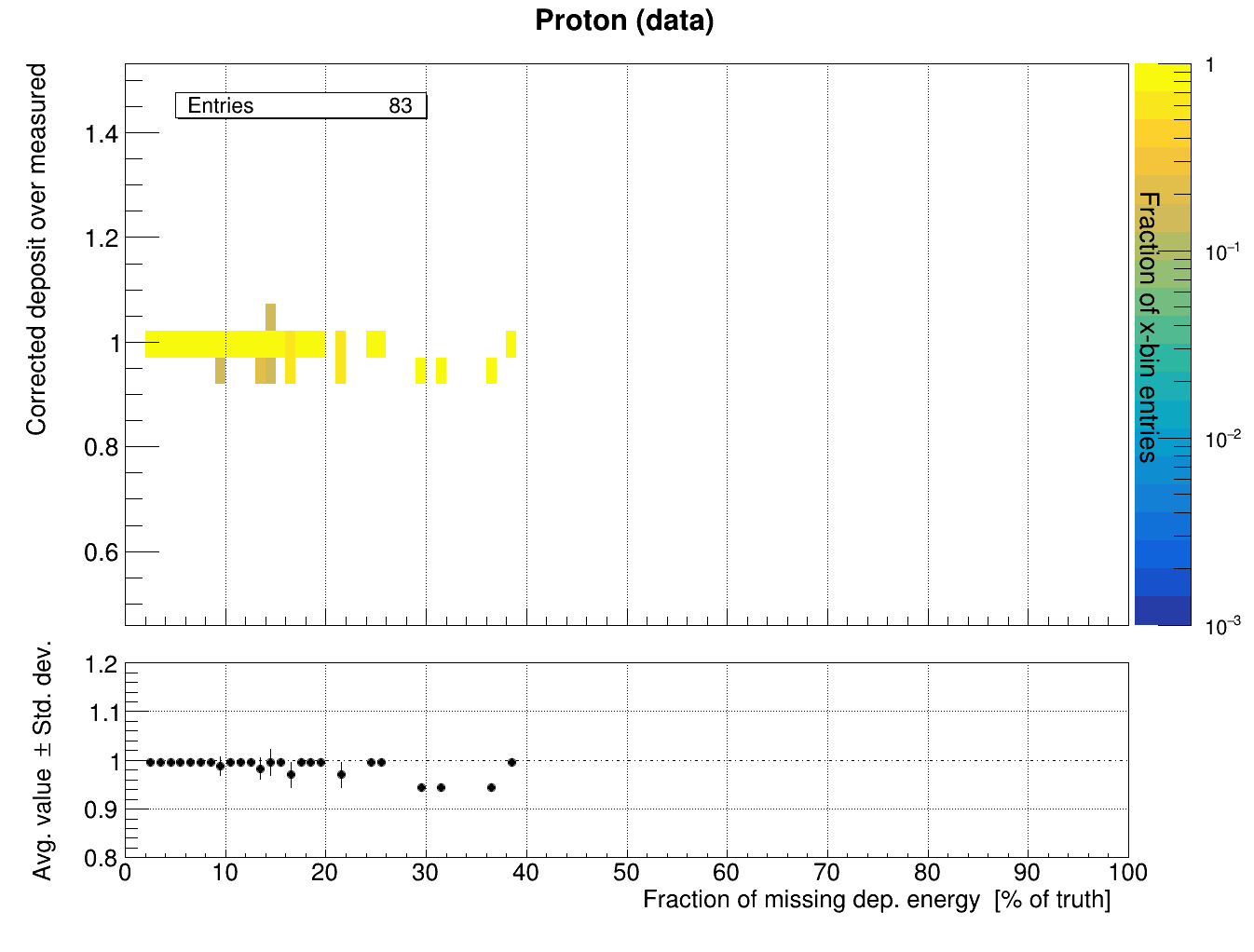}\\[3pt]
  \includegraphics[draft=false, width=0.48\textwidth]{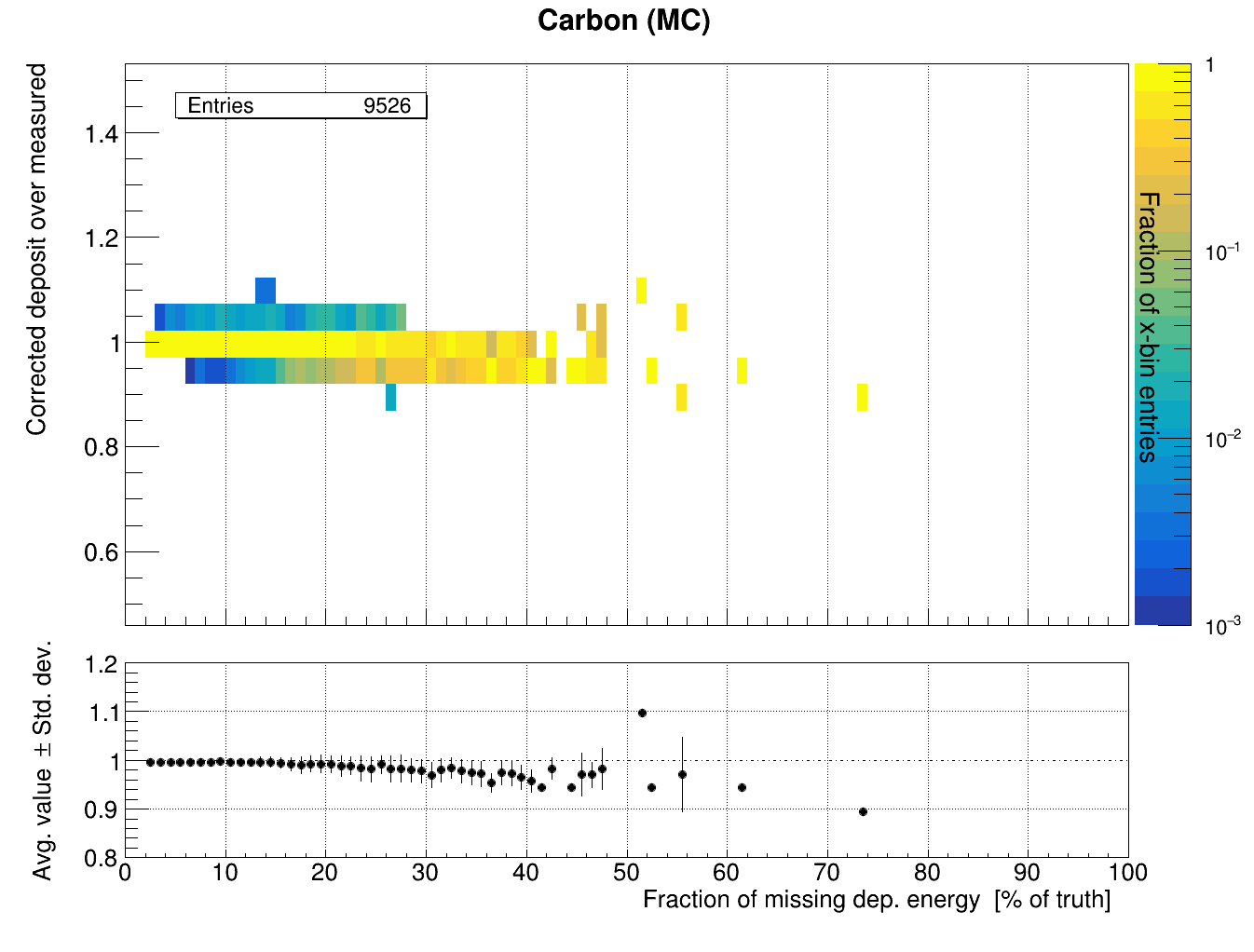}
  \includegraphics[draft=false, width=0.48\textwidth]{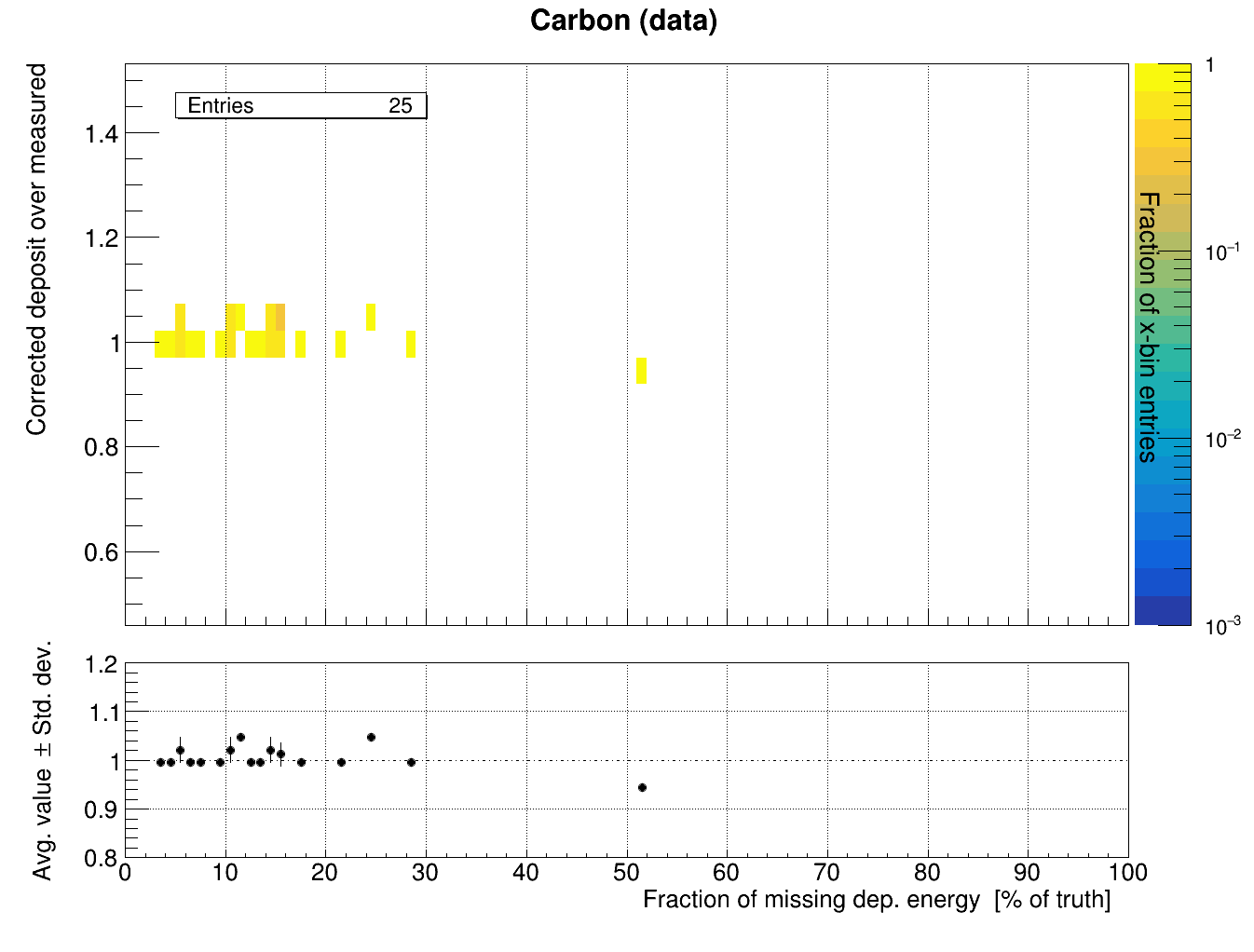}\\[3pt]
  \includegraphics[draft=false, width=0.48\textwidth]{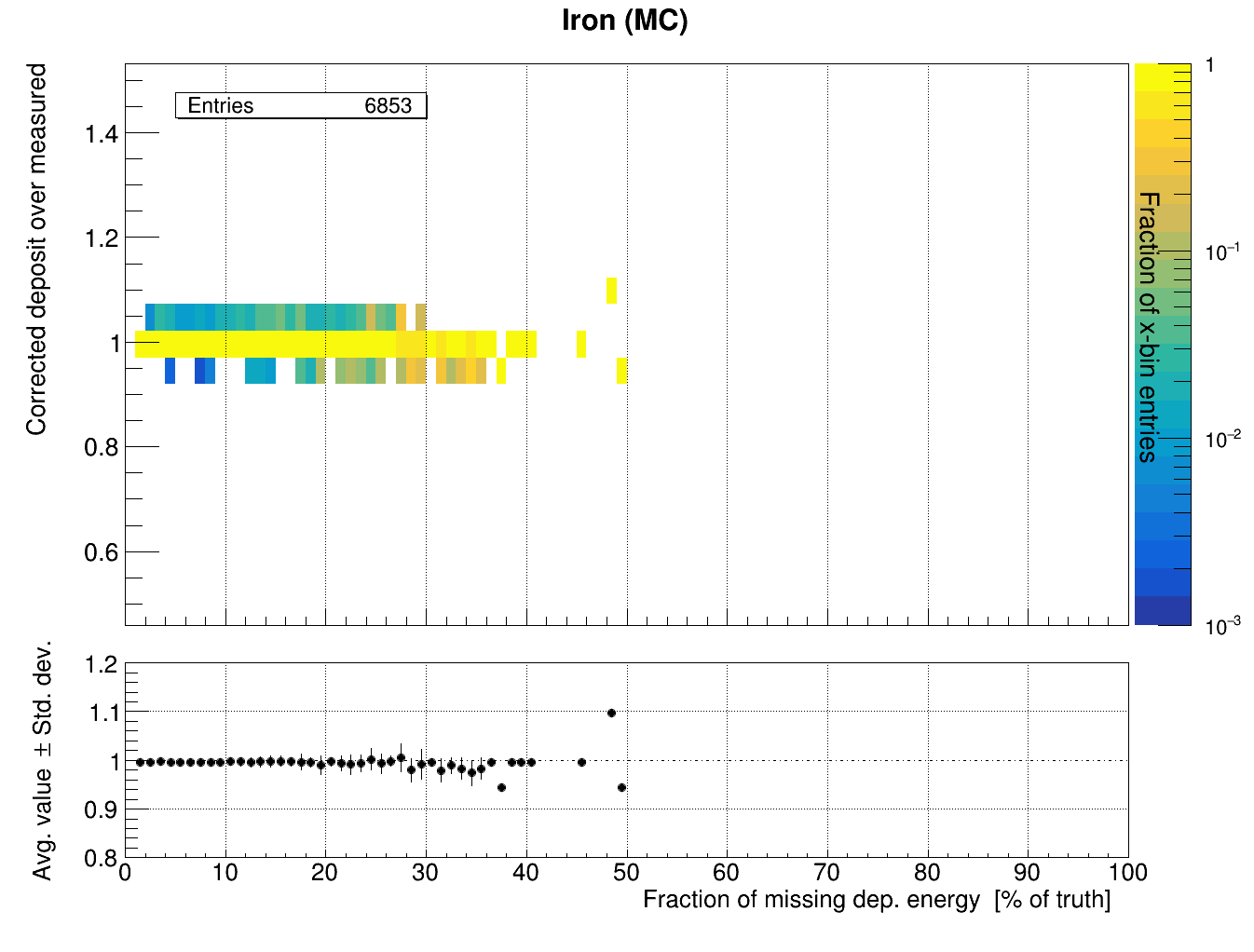}
  \includegraphics[draft=false, width=0.48\textwidth]{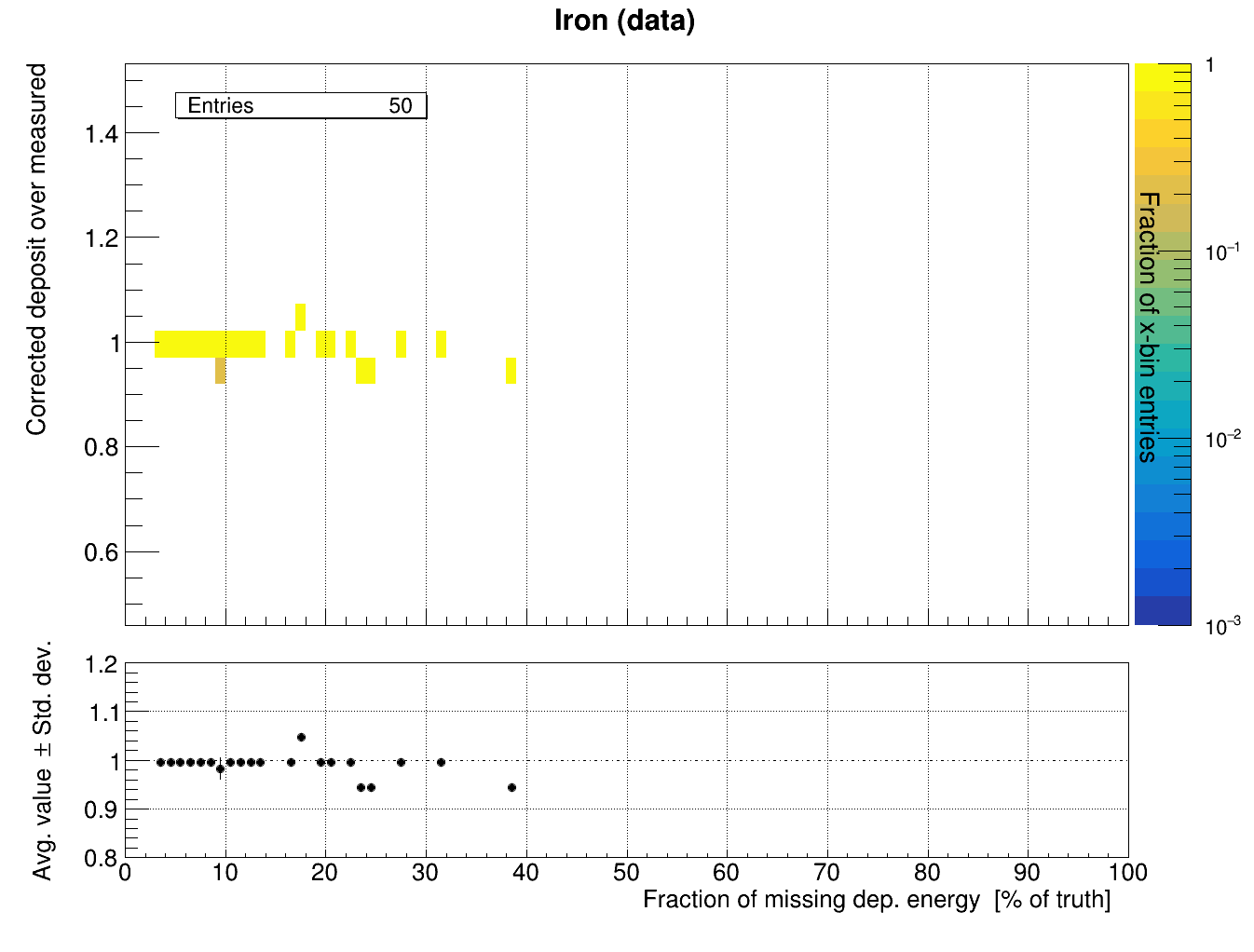}
  \caption{as per Figure \ref{fig:pseudo_simu}, but as a function of the fraction of energy lost in pseudo-saturation.\label{fig:pseudo_miss}}
\end{figure*}
\begin{figure*}[htb]
  \centering
  \includegraphics[draft=false, width=0.48\textwidth]{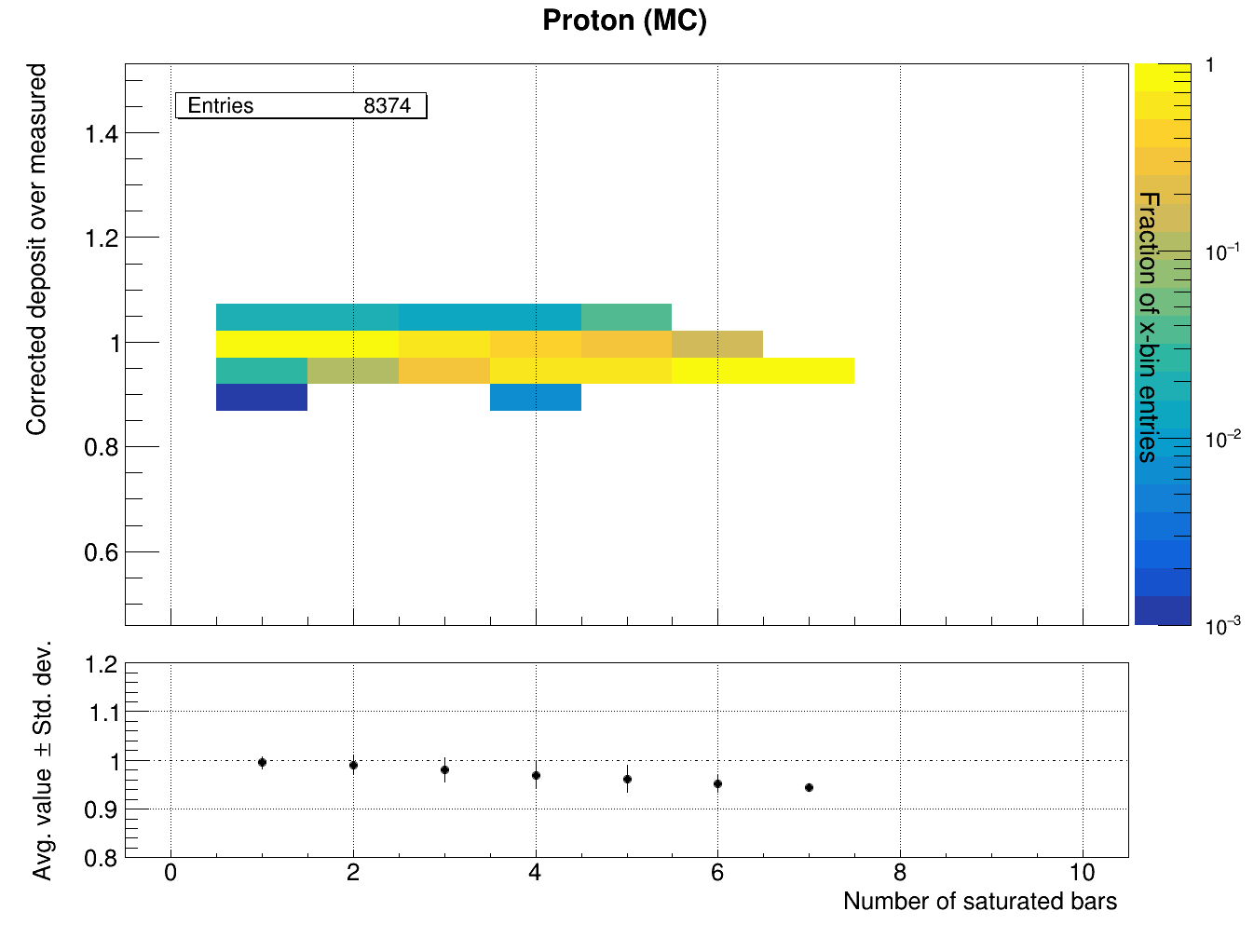}
  \includegraphics[draft=false, width=0.48\textwidth]{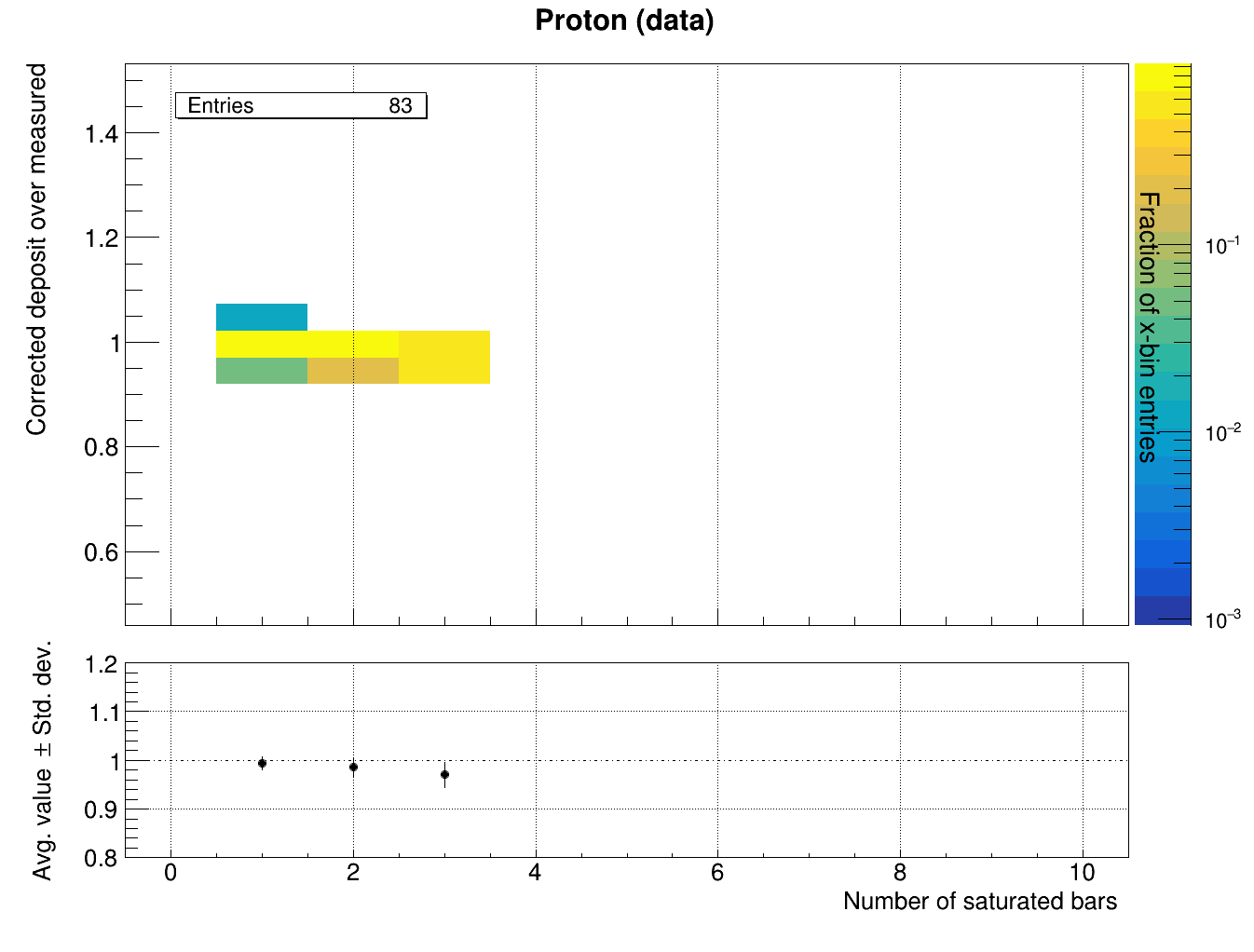}\\[3pt]
  \includegraphics[draft=false, width=0.48\textwidth]{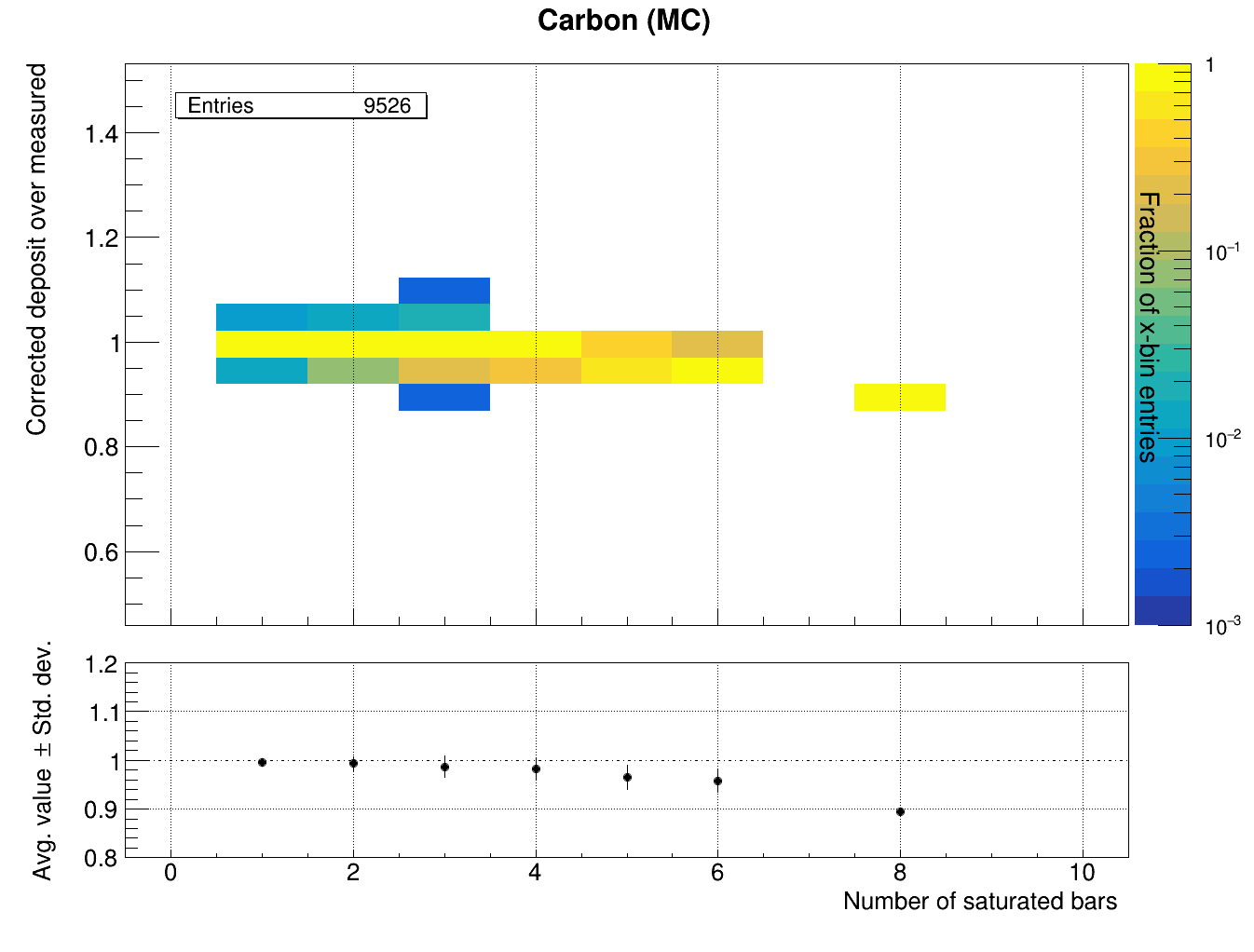}
  \includegraphics[draft=false, width=0.48\textwidth]{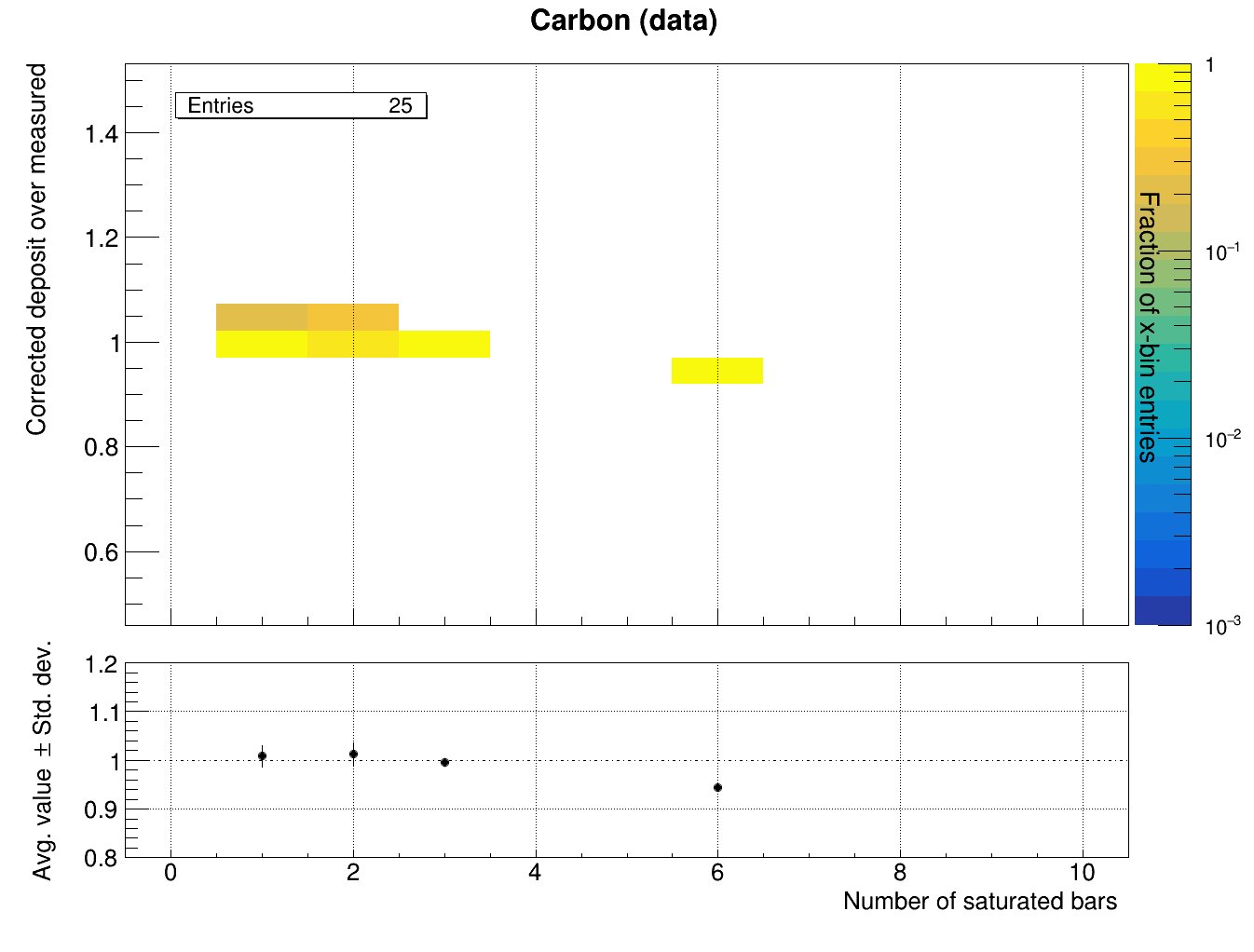}\\[3pt]
  \includegraphics[draft=false, width=0.48\textwidth]{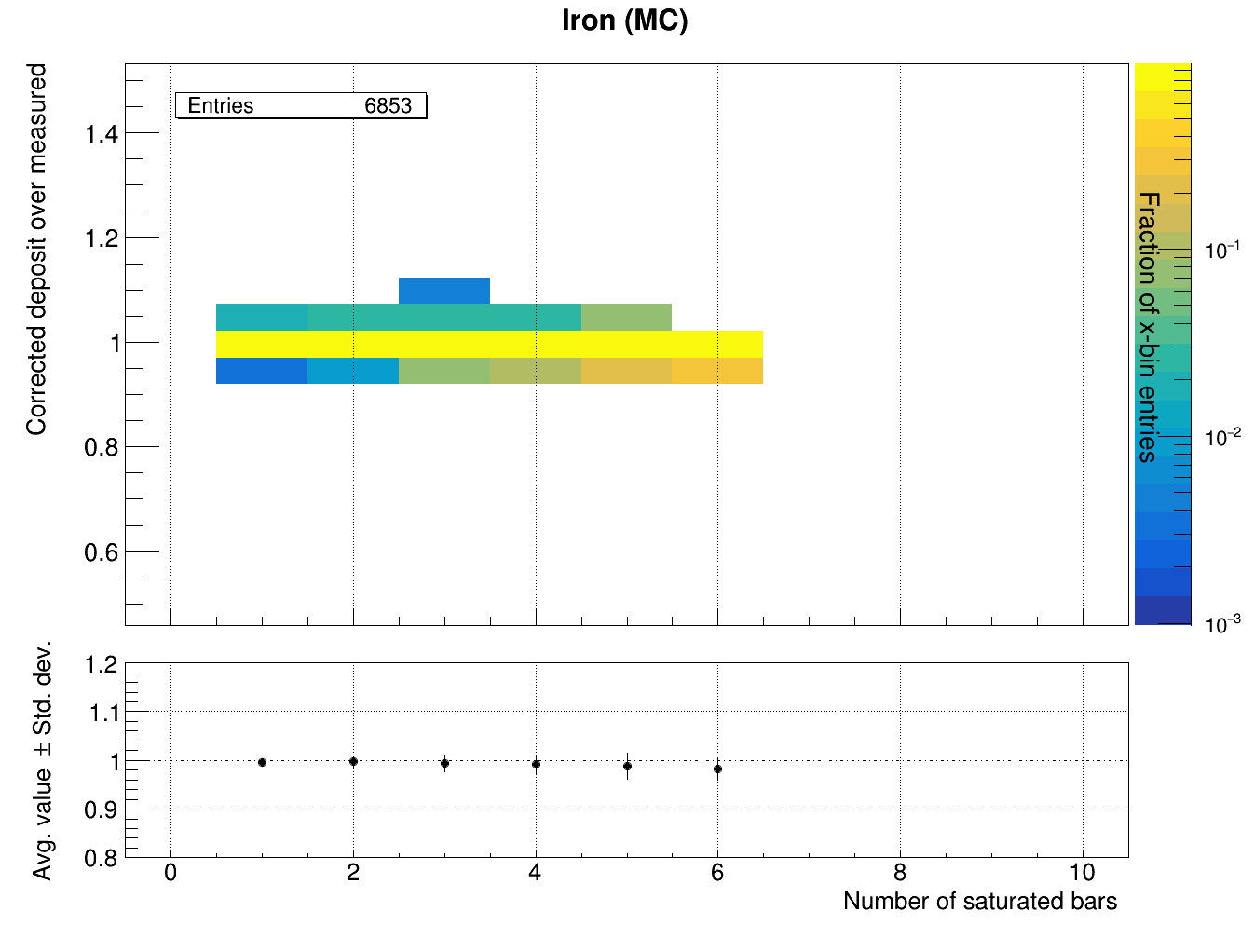}
  \includegraphics[draft=false, width=0.48\textwidth]{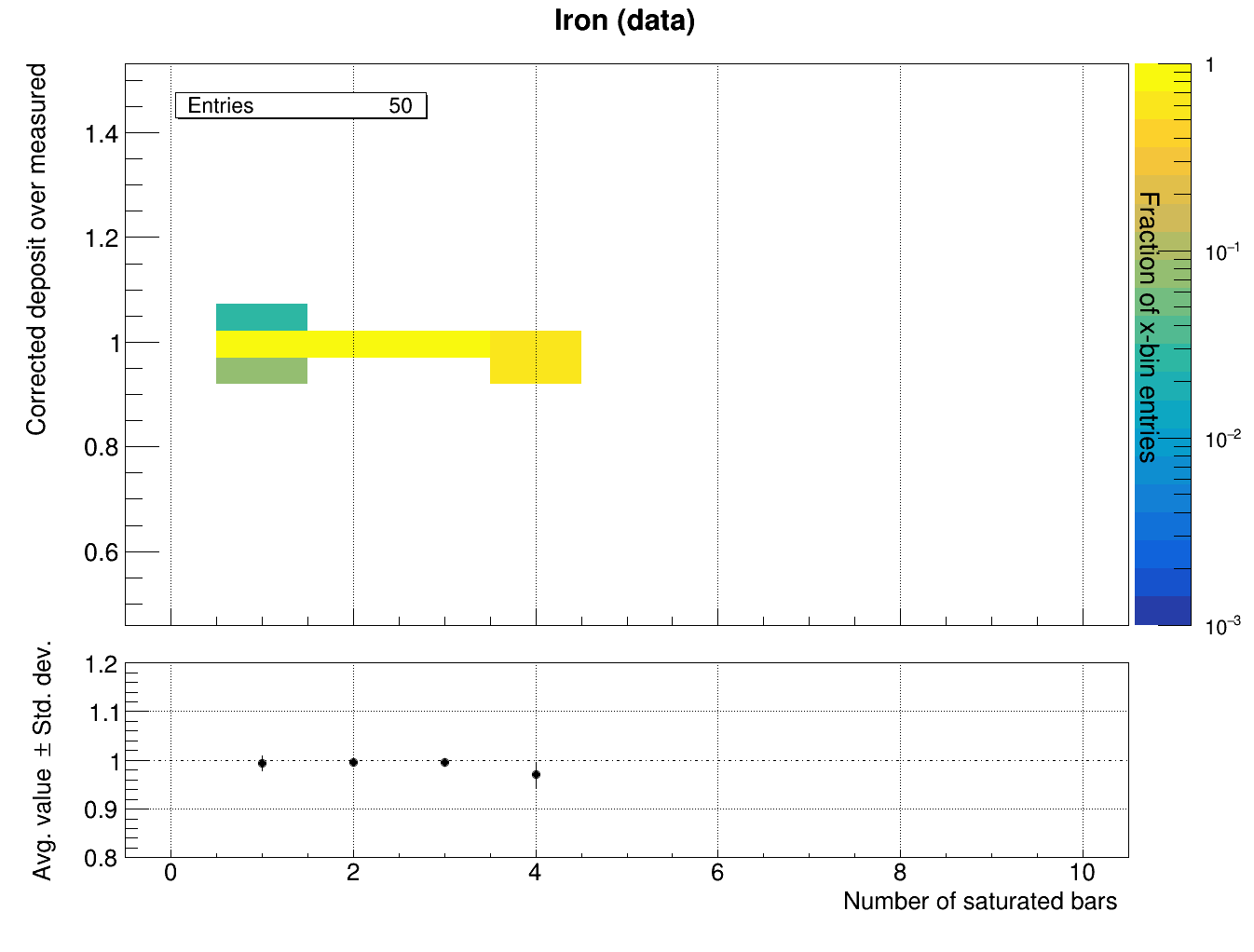}
  \caption{as per Figure \ref{fig:pseudo_simu}, but as a function of the number of pseudo-saturated bars.\label{fig:pseudo_nsat}}
\end{figure*}

\section{Conclusion and discussion}\label{sec:conclusion}

An ML method is developed to correct the deposited energies in the DAMPE calorimeter, in the presence of saturation.
The issue at hand provokes the partial loss of the original energy deposits, due to the inability of the readout electronics to process signals above their maximum capacity.
The use of a CNN makes it possible to analyse the BGO image with saturation and recover the original total deposited energy.

Prior to this work, two other corrections were developed to tackle the saturation in DAMPE calorimeter.
The work in \cite{DAMPE_saturation_analytical} provides an analytical method to infer the energy lost in saturation, using the information from the adjacent bars; the main limitation of this approach is the inability of application where several contiguous saturated bars are found.
On the other hand, the work in \cite{DAMPE_saturation_ML_2022} uses an ML technique to achieve a higher precision, but the final correction relies on two different CNNs to recover the energy lost respectively in the middle layers or in the last one.
The two models also need specific trainings for different ions.
Overall, both precedent works succeed in their aim, but they require specific tunings for different ions and start losing accuracy above $\sim$500~TeV of incident energy, posing issues for the flux measurements at the PeV scale.
In this work, the focus moved to the prediction of an overall correction factor for the total deposited energy, instead of reconstructing the single pixels of the BGO image.
This novel approach facilitates the model learning, achieving an improved precision.
In addition, the introduction of different ion species and higher incident energies in the model training results in an improved accuracy, up to several PeVs.
Both the improvements in maximum energy and generalisation supersede previous techniques, and can significantly help in probing the flux of CR ions at the PeV scale, where the knee region---i.e., a change in the all-particle spectral index---is expected to be.
On the other hand, highly saturated events, with more than 90\% of the original deposit lost, still require a deeper understanding to eliminate the residual bias.

The comparison between pseudo-saturated MC and data events shows a consistent behavior of the saturation correction on simulated and real events.

\section*{Acknowledgements}

The DAMPE mission was funded by the strategic priority science and technology projects in space science of Chinese Academy of Sciences.
In Europe the activities and data analysis are supported by the Swiss National Science Foundation (SNSF), Switzerland (grant No. 204253), the National Institute for Nuclear Physics (INFN), Italy, the European Research Council (ERC) under the European Union's Horizon 2020 research and innovation program, and the Swiss State Secretariat for Education, Research and Innovation (SERI).

\clearpage

\appendix

\section{Events re-weighting} \label{app:events_reweighting}

MC simulations generate events with a spectrum $\propto E_\mathrm{inc}^{-1}$, depending on the incident particle energy $E_\mathrm{inc}$.
Experimental measurements suggest that a spectral index of $-$2.7 is more appropriate.
To modify the spectral index of simulated events to $\gamma$, a weight $w$ can be associated with each entry based on the definition
\begin{equation} \label{eq:weights}
  w = \frac{E_\mathrm{inc}^{(1 - \gamma)}}{N_\mathrm{tot}\ M_\mathrm{samples}}\text{ ,}
\end{equation}
with $N_\mathrm{tot}$ the total number of events, and $M_\mathrm{samples}$ the number of different samples populating the same energy range to which $E_\mathrm{inc}$ belongs.
Figure \ref{fig:reweight_prim} shows the distributions of proton, carbon and iron events used in this work before and after re-weighting to an $E^{-2.7}$ spectrum.

\begin{figure*}[htb]
  \centering
  \includegraphics[draft=false, width=0.48\linewidth]{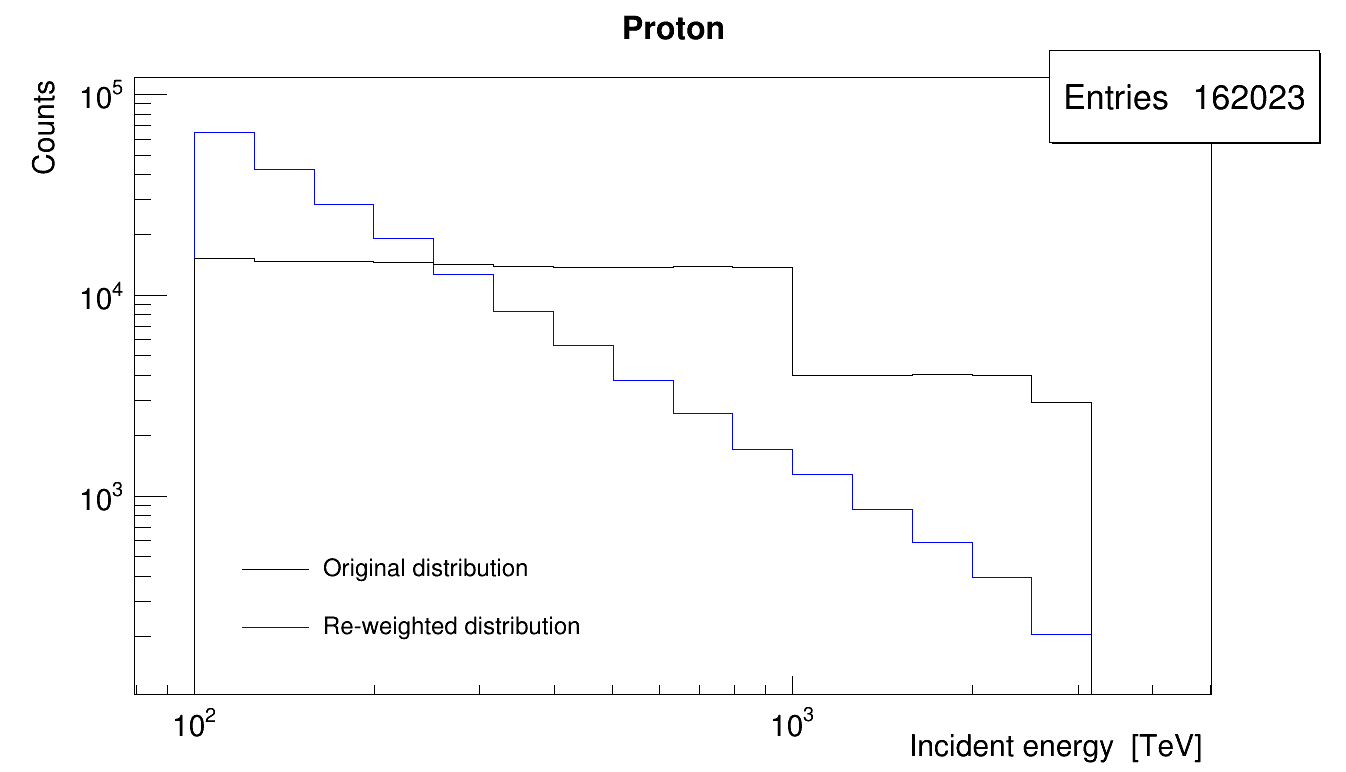}
  \includegraphics[draft=false, width=0.48\linewidth]{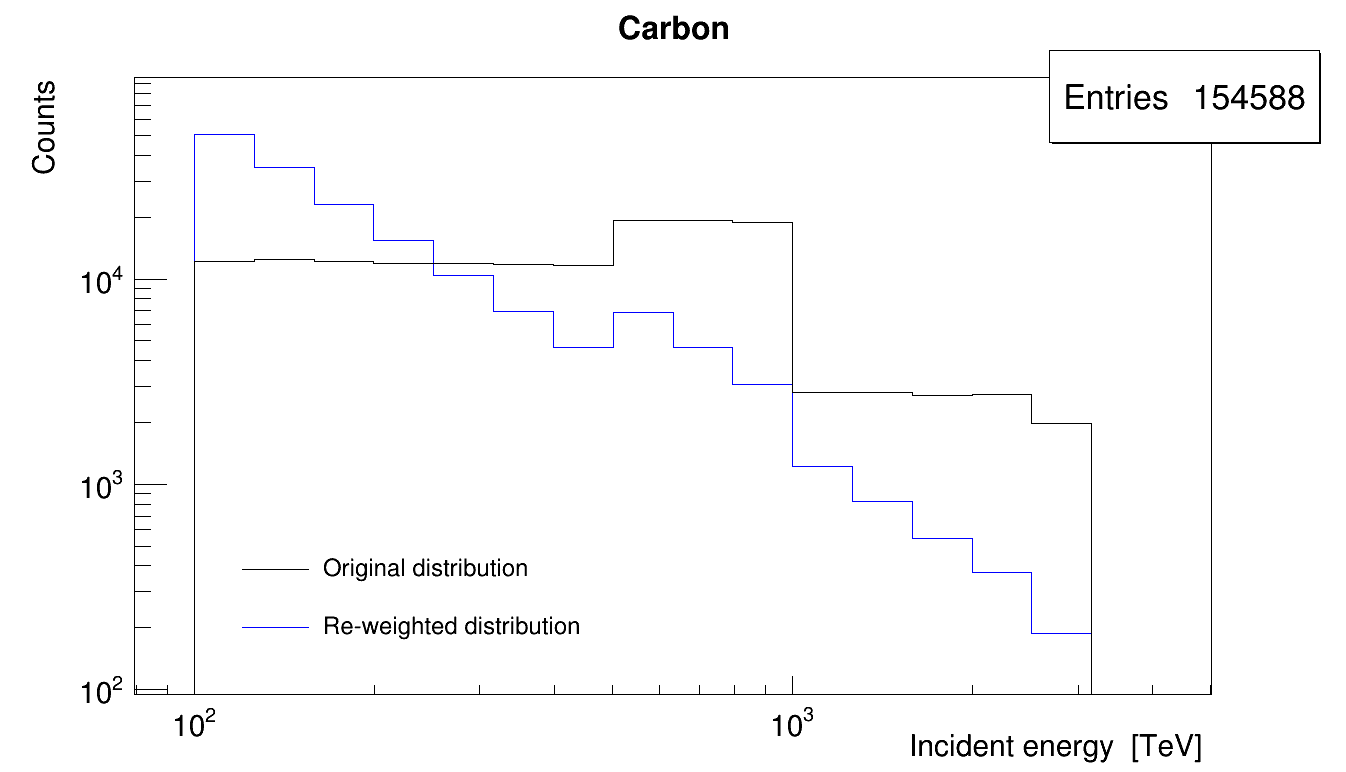}\\[3pt]
  \includegraphics[draft=false, width=0.48\linewidth]{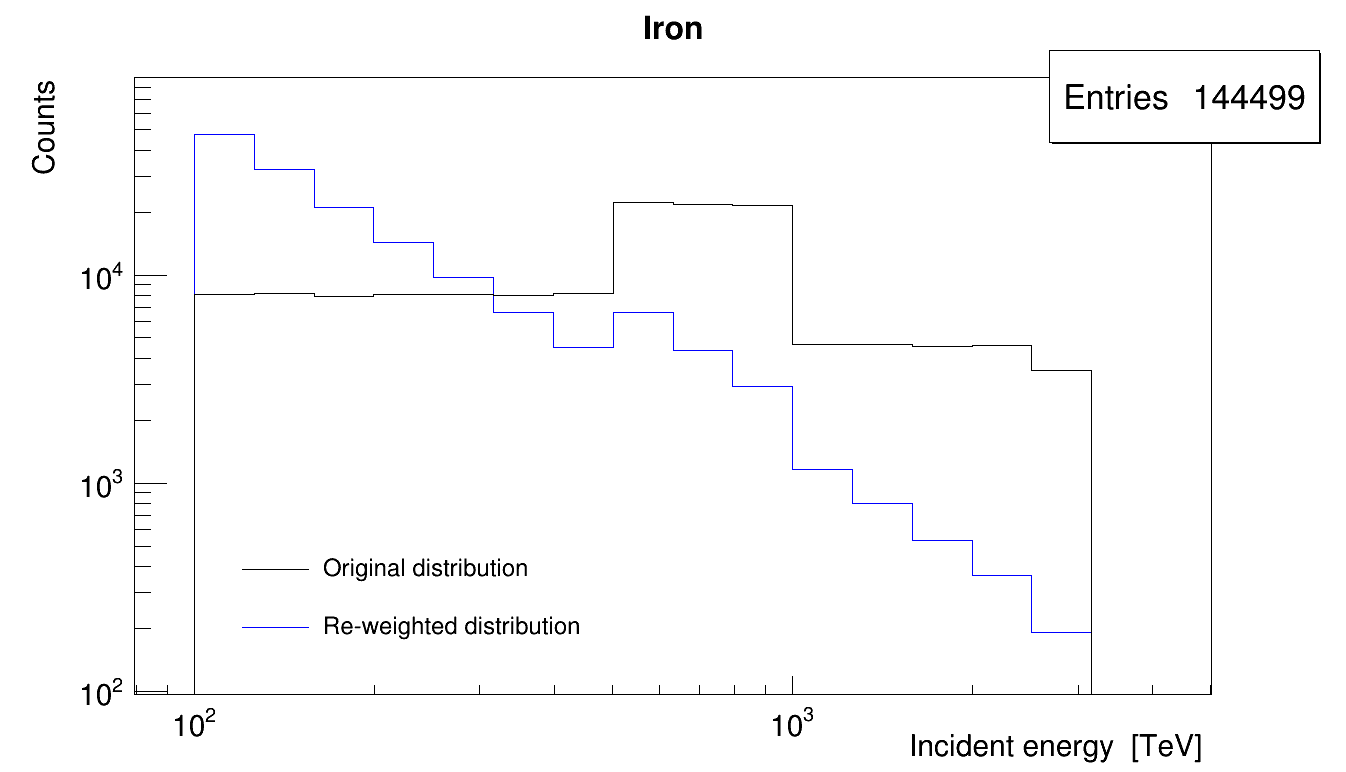}
  \caption{distribution of the primary energy, before and after re-weighting the entries to a simil-CR spectrum, proportional to $E_\mathrm{inc}^{-2.7}$. Three different plots are shown for the proton (top left), carbon (top right) and iron events (bottom). The distributions consider the full simulated samples after event selection.\label{fig:reweight_prim}}
\end{figure*}

\clearpage

\section{Fraction of saturated events} \label{app:sat_frac}

Generally, BGO bars of DAMPE calorimeter saturate more as the incident energy increases.
Figure \ref{fig:sat_frac} shows the fraction of saturated events found in MC simulations of proton, carbon and iron, as a function of the incident energy.

\begin{figure}[htb]
  \centering
  \includegraphics[draft=false, width=\ifoptiondraft{0.6\linewidth}{\linewidth}]{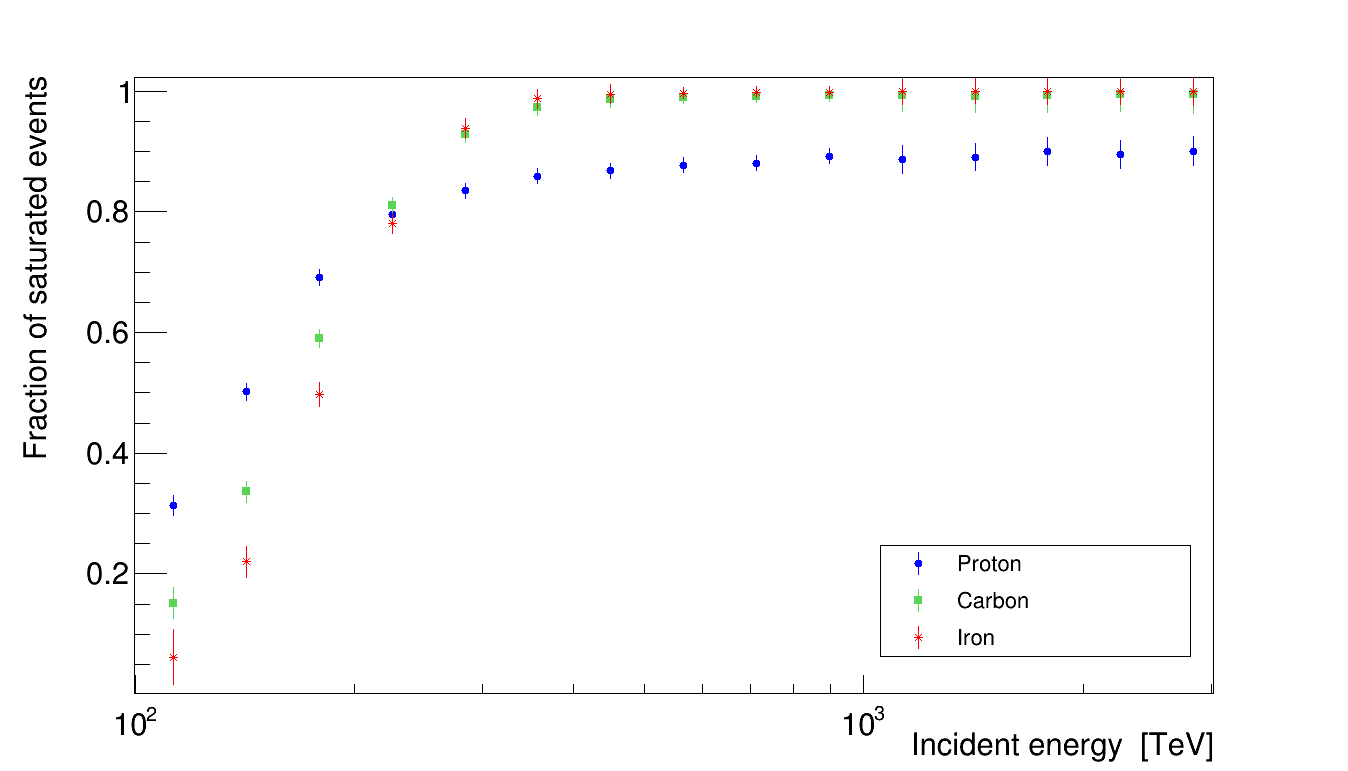}
  \caption{relative fraction of saturated events in MC simulations of protons, carbon and iron nuclei crossing DAMPE instrument. The fraction is plotted as a function of the true incident energy. The absolute total number of events can be found in Figure \ref{fig:reweight_prim}.\label{fig:sat_frac}}
\end{figure}

\section{PSD charge calculation}\label{app:psd_charge}

The DAMPE PSD can provide up to 4 independent measurements of the traversing particle charge.
From all the measured deposits in the scintillators of a sub-layer, the one closest to the particle track is selected for the average charge measurement.
For tracking purposes, the ML algorithm described in \cite{DAMPE_ML_tracking} is employed.
After locating the closest PSD hits to the candidate track, the deposits are corrected according to the traversed path in the scintillator material; hits with a path length shorter than 5~mm are discarded.
Having obtained the vector of independent charge measurements, a top-to-bottom progressive average is calculated.
A hit is permitted to differ from the partial average by 2.
When an inconsistent hit is found, the particle has likely interacted inelastically with the detector material, therefore the observed hit and all subsequent ones are rejected, and the average is finalised.

\nolinenumbers

\bibliographystyle{elsarticle-num}
\bibliography{bibliography.bib}

\end{document}